\documentclass[useAMS,usenatbib]{mn2e}

\newcommand{\etal}{\mbox{et\ al.\ }}

\usepackage{rotating}
\usepackage{url}

\title[\emph{Swift} follow-up of X-ray sources in the \emph{XMM} Slew Survey]
{\emph{Swift} follow-up of unidentified X-ray sources in the \emph{XMM-Newton} Slew Survey}

\author[Starling \etal 2010]
{R.L.C. Starling$^1$, P.A. Evans$^1$, A.M. Read$^1$, R.D. Saxton$^2$, P. Esquej$^1$, H. Krimm$^{3,4}$,\and
P.T. O'Brien$^1$, J.P. Osborne$^1$, S. Mateos$^1$, R. Warwick$^1$ and K. Wiersema$^1$ \\
$^1$Department of Physics and Astronomy, University of Leicester, University Road, Leicester LE1 7RH, UK.\\$^2${\it XMM-Newton} SOC, ESAC, Apartado 78, 28691 Villanueva de la Ca\~nada, Madrid, Spain. \\$^3$Universities Space Research Association, Columbia, MD 20144, USA.\\$^4$NASA/Goddard Space Flight Center, Greenbelt, MD 20771, USA.}

\begin{document}
\date{Accepted 2010 November 11. Received 2010 November 11; in original form 2010 July 22.}

\pagerange{\pageref{firstpage}--\pageref{lastpage}} \pubyear{2010}

\maketitle

\label{firstpage}


\begin{abstract}
We present deep {\it Swift} follow-up observations of a sample of 94 unidentified X-ray sources from the {\it XMM-Newton} Slew Survey. The X-ray Telescope on-board {\it Swift} detected 29\% of the sample sources; the flux limits for undetected sources suggests the bulk of the Slew Survey sources are drawn from one or more transient populations. We report revised X-ray positions for the XRT-detected sources, with typical uncertainties of 2.9$''$, reducing the number of catalogued optical matches to just a single source in most cases. We characterise the sources detected by {\it Swift} through their X-ray spectra and variability and via UVOT photometry and using catalogued nIR, optical and radio observations of potential counterparts. 
Six sources can be associated with known objects and 8 sources may be associated with unidentified {\it ROSAT} sources within the 3$\sigma$ error radii of our revised X-ray positions. We find 10 of the 30 XRT- and/or BAT-detected sources are clearly stellar in nature, including one periodic variable star and 2 high proper motion stars. For 11 sources we propose an AGN classification, among which 4 are detected in hard X-rays and 3 have redshifts spanning $z = 0.2 - 0.9$ obtained from the literature or from optical spectroscopy presented here. A further 3 sources are suspected AGN and 1 is a candidate Galactic hard X-ray flash, while 5 sources remain unclassified.
The 67 Slew Survey sources we do not detect with {\it Swift} XRT or BAT are studied via their characteristics in the Slew Survey observations and by comparison with the XRT and BAT detected population. We suggest that these are mostly if not all extragalactic, though unlikely to be highly absorbed sources in the X-rays such as Compton thick AGN. A large number of these are highly variable soft X-ray (0.2--2 keV) sources and a smaller number are highly variable hard (2--12 keV) sources. A small fraction of mainly hard-band Slew Survey detections may be spurious. This follow-up programme brings us a step further to completing the identifications of a substantial sample of {\it XMM-Newton} Slew Survey sources, important for understanding the nature of the transient sky and allowing flux-limited samples to be constructed.
\end{abstract}

\begin{keywords}
Surveys --- X-rays
\end{keywords}

\section{Introduction}
The {\it XMM-Newton} Slew Survey \citep{Saxton} performed with the pn channel of the European Photon Imaging Camera (EPIC) \citep[][see also \citealt{Jansen}]{Struder} is proving to be a useful resource for the discovery of
bright new X-ray sources. The Slew Survey makes use of data taken while the satellite is manouevering between pointed observations, reaching five to ten times deeper in flux than all other all-sky spatially-resolved surveys in
the 2--12\,keV band. It also reaches comparable sensitivity to the {\it ROSAT} PSPC All-Sky
Survey  \citep[RASS,][]{Voges99,Voges00} in the 0.2--2\,keV band. The latest release of the clean slew catalogue (XMMSL1 - delta4) contains 11425 sources detected over 28000 deg$^2$ of which 72\% are previously known in X-rays or have plausible counterparts
from other wavebands. Several interesting transients have been discovered including novae \citep{Read08nova,Read09}, tidal disruption candidates \citep{Esquej} and flare stars \citep[e.g.][]{Read08,Saxton08}. However, a quarter of the {\it XMM-Newton} Slew Survey sources are relatively bright yet appear to have no previous, catalogued X-ray detections. Cross-correlation with the RASS showed that of order 50\% of the Slew Survey point-like sources do not have RASS counterparts. 

Potential explanations for the lack of previous X-ray detections of these Slew Survey sources include transient or highly variable X-ray
behaviour (perhaps such sources are seen in a `high' state during the {\it XMM-Newton} observations) or hard X-ray spectra (meaning that most of the counts fall outside the {\it ROSAT} energy range). Further possibilities include an inaccurate Slew Survey position: the 1$\sigma$ position error is 8$''$ but this has a long tail (see
Figure 6 of \citealt{Saxton}); or spurious detections: $\sim$4\% of the sources in the clean Slew Survey catalogue are expected to be
spurious from statistical considerations \citep{Saxton}.
It is important to try and complete the identifications of the {\it XMM-Newton} Slew Survey catalogue, so as to allow
flux-limited samples to be drawn from the Survey and to develop a fuller picture of the X-ray transient source population.

Here we present follow-up observations of a sample of unidentified {\it XMM-Newton} Slew Survey X-ray sources with the {\it Swift} satellite \citep{Gehrels}. We attempt to classify the sources by obtaining more accurate localisations with the {\it Swift} X-Ray Telescope, measuring any X-ray variability and where possible identifying the broadband spectral properties using all instruments on-board {\it Swift} and information from published optical, near infrared (nIR) and radio catalogues. The main body of the paper details the sample, our analyses and general results. Discussion of individual sources is given in the Appendix.

\section{Sample selection} \label{sample}
We selected our sample from the XMMSL1 catalogue (2007 August) to include sources which, in either the full (0.2$-$12\,keV), hard (2$-$12\,keV) or soft (0.2$-$2\,keV) bands: (a) were detected with likelihood $\ge$\,10, (b) were detected with $\ge$\,4 counts, (c) had a low value of fitted source extent (best band extent $\le$\,10 pixels), and (d) were not consistent with any known source in a multiple-catalogue search (including SIMBAD, NED and RASS to within 30$''$, see Table 6 of \citealt{Saxton} for the complete list).
This resulted in 97 sources, of which 94 have {\it Swift} pointed observations and are presented here. Full-band (0.2--12 keV) X-ray fluxes for the sample, as given in the XMMSL1 catalogue, range from $\sim$(2--30)$\times$10$^{-12}$ erg cm$^{-2}$ s$^{-1}$ and the sources are distributed apparently randomly across the sky (Fig. \ref{aitoff}).

\section{\emph{Swift} observations}
We observed the 94 sources defined in Section \ref{sample} with all instruments on-board {\it Swift} simultaneously: the wide field of view Burst Alert Telescope \citep[BAT,][]{Barthelmy} operating in the energy range 15--150 keV, and the narrow-field instruments - the X-Ray Telescope \citep[XRT,][]{Burrows} and the UltraViolet-Optical Telescope \citep[UVOT,][]{Roming}. The observations with XRT (Table \ref{tab:obs}) were designed to be performed in Photon Counting (PC) mode and to have a minimum exposure time of 1.8\,ks, to obtain an improved X-ray position and some spectral information. These observations were performed as `fill-in' targets which can be overridden when {\it Swift} slews to higher priority targets such as gamma-ray bursts, resulting in exposure times varying from 240\,s up to 10220\,s. The total exposure time per source may have been continuously accumulated or be spread over a number of months, from 2006 August to 2009 December.
Where possible, observations with the UVOT have been carried out using the $b$ filter to optimise UVOT-enhanced X-ray position determination (described in Section \ref{positions}). BAT data from the pointed observations have been combined with data from the BAT 58-month Survey (Baumgartner et al. 2010, in preparation; see also \citealt{Tueller}) to increase detection likelihood.

\begin{table*}
\caption{{\it Swift} XRT observations of unidenitified {\it XMM-Newton} Slew Survey sources. XRT count rates are PC mode 0.3--10\,keV with 1$\sigma$ errors, or 3$\sigma$ upper limits where no source was detected at $\ge$3$\sigma$ significance (in which case no error is given). The {\it XMM-Newton} Slew Survey bands are described in Section \ref{sample}.}
\label{tab:obs}
\begin{center}
\begin{tabular}{l c c c l l l l}
Source & \multicolumn{3}{c}{Slew count rate (ct s$^{-1}$)}&{\it Swift} & date$-$obs start, end & T$_{\rm exp}$  & XRT ct rate/U. lim. \\
 & Full& Hard& Soft & obsID &  &  (ks) &  (ct s$^{-1}$)\\ \hline
  XMMSL1\,J002202.9+254004  &      2.6$\pm$0.7&      -&      2.0$\pm$0.5    & 00037850 & 2008 Aug, 2009 Jun&2.38 & 0.138$\pm$0.005 \\
  XMMSL1\,J003023.0+515845  &      2.4$\pm$0.8&      2.7$\pm$0.8&    -   & 00037851 & 2008 May, Jun & 3.69 & $\le$1.57$\times$10$^{-3}$ \\
  XMMSL1\,J004712.3+353738  &     0.8$\pm$0.3&      1.0$\pm$0.4&      -    & 00035862 & 2007 Jun& 1.96 & $\le$3.30$\times$10$^{-3}$ \\
  XMMSL1\,J010654.8+802740  &      1.8$\pm$0.4&      -&      1.6$\pm$0.4    & 00037861  & 2008 Sep & 2.72 & 0.016$\pm$0.003 \\
  XMMSL1\,J011407.7+124648  &      1.9$\pm$0.6&      -&      1.7$\pm$0.3    & 00037857  & 2008 Aug, 2009 Jun & 2.12 &$\le$2.23$\times$10$^{-3}$ \\
 XMMSL1\,J012240.2$-$570859 &      1.6$\pm$0.5&      -&      1.3$\pm$0.4   & 00035821  & 2007 Mar, Dec & 2.86 & 0.043$\pm$0.004 \\
   XMMSL1\,J014957.3+365200 &      1.7$\pm$0.8&      2.0$\pm$1.0&      -     & 00035810  & 2008 Feb & 3.27 & $\le$2.14$\times$10$^{-3}$ \\
 XMMSL1\,J025808.2$-$651845 &      1.8$\pm$0.6&      -&     1.0$\pm$0.4   & 00035808  & 2006 Nov, 2007 Dec & 5.33 & $\le$1.64$\times$10$^{-3}$\\
 XMMSL1\,J030006.6$-$381617 &      1.4$\pm$0.3&      -&      1.0$\pm$0.3   & 00035842  & 2006 Nov & 2.03 & 0.021$\pm$0.003 \\
 XMMSL1\,J033952.5$-$651256 &      1.3$\pm$0.4&     0.8$\pm$0.4&      -   & 00037873  & 2008 Oct & 6.36 & $\le$1.02$\times$10$^{-3}$ \\
XMMSL1\,J034923.7$-$433330  &      1.2$\pm$0.3&      -&     1.0$\pm$0.3  & 00037877  & 2008 Oct & 2.36 & $\le$4.55$\times$10$^{-3}$ \\
XMMSL1\,J035115.5$-$434049  &      1.6$\pm$0.5&      -&      1.1$\pm$0.4  & 00037862  & 2008 Oct, Dec & 2.51 & $\le$2.78$\times$10$^{-3}$ \\
  XMMSL1\,J043707.5+112538  &      1.5$\pm$0.7&      -&      -    & 00035828   &  2007 Mar & 2.33 & $\le$2.50$\times$10$^{-3}$ \\
 XMMSL1\,J044357.4$-$364413 &      2.4$\pm$0.6&      -&      1.5$\pm$0.5   & 00035791  & 2006 Aug & 2.24 & $\le$3.31$\times$10$^{-3}$ \\
 XMMSL1\,J045937.1$-$153256 &     1.0$\pm$0.4&      1.1$\pm$0.5&      -   & 00035858  & 2006 Aug, 2008 Apr & 4.43 & $\le$1.57$\times$10$^{-3}$ \\
XMMSL1\,J045949.8$-$573514  &      1.3$\pm$0.5&      1.4$\pm$0.5&      -  & 00035849  & 2006 Aug & 2.18 & $\le$4.02$\times$10$^{-3}$ \\
XMMSL1\,J050801.1$-$284113  &      1.7$\pm$0.5&      -&      1.3$\pm$0.4  & 00035815  & 2006 Aug, 2008 Feb & 2.64 & $\le$2.20$\times$10$^{-3}$\\
  XMMSL1\,J050824.5+220834  &      2.1$\pm$0.6&      -&      -    & 00035795   & 2006 Sep & 3.05 & $\le$2.67$\times$10$^{-3}$  \\
 XMMSL1\,J060339.9$-$294302 &      3.5$\pm$1.0&      -&      1.4$\pm$0.6   & 00035786  & 2006 Oct & 2.54 & $\le$2.29$\times$10$^{-3}$\\
  XMMSL1\,J060730.8+691832  &      1.8$\pm$0.8&      -&      1.2$\pm$0.5    & 00035805  & 2006 Aug & 2.34 & $\le$3.16$\times$10$^{-3}$ \\
  XMMSL1\,J063950.7+093634  &      1.4$\pm$0.5&      -&      1.2$\pm$0.3    & 00037869  & 2008 Aug & 1.86 & $\le$3.13$\times$10$^{-3}$ \\ 
XMMSL1\,J064041.6$-$582308  &      1.2$\pm$0.3&      -&     0.4$\pm$0.2  & 00037878  & 2008 Nov & 2.10 & $\le$2.76$\times$10$^{-3}$ \\
XMMSL1\,J064109.2$-$565542  &      1.4$\pm$0.4&      -&      1.0$\pm$0.3  & 00037870  & 2008 Oct & 5.56 & 0.039$\pm$0.003 \\
   XMMSL1\,J064849.0+394715 &      1.1$\pm$0.3&      -&      1.0$\pm$0.3    & 00035845  & 2008 Jan & 5.89 & $\le$9.86$\times$10$^{-4}$ \\
   XMMSL1\,J065525.2+370815 &      2.5$\pm$0.4&     0.5$\pm$0.2&      1.8$\pm$0.3 & 00035789  & 2006 Oct, 2008 Jan & 3.25 & 0.027$\pm$0.006 \\
   XMMSL1\,J070846.2+554905 &      1.5$\pm$0.6&      -&      -     & 00035787  & 2006 Oct, 2007 Jan& 3.84 & 0.067$\pm$0.004 \\
                            &      3.1$\pm$0.6&     0.6$\pm$0.2&      2.3$\pm$0.4   & & & & \\
  XMMSL1\,J071111.8+280314  &      2.6$\pm$0.8&      -&      1.2$\pm$0.5     & 00037849  & 2009 May & 1.02 & $\le$6.84$\times$10$^{-3}$\\
XMMSL1\,J075818.9$-$062723  &      -&      -&      2.6$\pm$1.2  & 00035835  & 2007 Feb, Apr& 5.35 & 0.020$\pm$0.002 \\
XMMSL1\,J080849.0$-$383803  &      1.4$\pm$0.4&      -&      1.0$\pm$0.2  & 00037871  & 2008 Jun, Jul & 1.78 & 0.063$\pm$0.006 \\
XMMSL1\,J082730.0$-$672401  &      3.9$\pm$0.6&     0.6$\pm$0.2&      3.2$\pm$0.6  & 00037847  & 2008 Dec, 2009 Jan& 2.20 & $\le$3.17$\times$10$^{-3}$ \\
  XMMSL1\,J083704.0+193951  &      1.8$\pm$0.8&      -&      1.4$\pm$0.6  & 00035807  & 2007 May & 5.92 & $\le$1.37$\times$10$^{-3}$ \\  
XMMSL1\,J084756.4$-$532755  &      1.6$\pm$0.6&      -&      1.2$\pm$0.5  & 00035824  & 2007 Feb & 1.66 & $\le$3.90$\times$10$^{-3}$ \\
XMMSL1\,J084945.3$-$413706  &      1.7$\pm$0.4&      -&      1.1$\pm$0.3  & 00035811  & 2007 Apr, May & 4.34 & $\le$2.32$\times$10$^{-3}$ \\
 XMMSL1\,J085036.8+044354   &      1.9$\pm$0.5&      1.1$\pm$0.4&     0.7$\pm$0.3 & 00037860  & 2009 Jan & 0.24 & $\le$3.89$\times$10$^{-2}$ \\
XMMSL1\,J085155.6$-$570352  &      1.5$\pm$0.4&      -&      1.2$\pm$0.3 & 00035838  & 2007 Jan & 2.29 & $\le$3.83$\times$10$^{-3}$ \\
 XMMSL1\,J085216.3+283657   &      1.5$\pm$0.5&      -&      1.1$\pm$0.4   & 00037866  & 2009 Mar & 2.39 & $\le$2.71$\times$10$^{-3}$ \\
XMMSL1\,J090822.3$-$643749  &      9.9$\pm$0.9&      1.7$\pm$0.4&      7.0$\pm$0.8  & 00035781  & 2006 Dec & 3.05 & $\le$2.12$\times$10$^{-3}$ \\
  XMMSL1\,J092118.6+015302  &      2.3$\pm$0.6&      -&      1.4$\pm$0.5   & 00035793  & 2007 Oct & 2.76 & $\le$2.10$\times$10$^{-3}$ \\
 XMMSL1\,J093738.4$-$654445 &      1.4$\pm$0.6&      1.6$\pm$0.6&      -   & 00035847  & 2007 Jan, Apr & 3.26 & $\le$2.14$\times$10$^{-3}$ \\
  XMMSL1\,J094156.1+163246  &      1.7$\pm$0.5&      1.8$\pm$0.5&      -    & 00035813  & 2007 Jun, 2008 Jun & 3.56 & $\le$2.08$\times$10$^{-3}$ \\
XMMSL1\,J094551.3$-$194352  &      1.6$\pm$0.4&      -&     1.0$\pm$0.3 & 00037863  & 2009 Feb & 10.22 & 0.049$\pm$0.002 \\
  XMMSL1\,J095336.4+161231  &      1.5$\pm$0.4&      -&      1.3$\pm$0.4    & 00035837  & 2007 Jun, 2008 Jun & 2.46 & 0.017$\pm$0.003\\
  XMMSL1\,J100011.5+553035  &      1.3$\pm$0.4&      -&     0.6$\pm$0.3   & 00037874  & 2009 Feb & 2.27 & $\le$2.85$\times$10$^{-3}$ \\
XMMSL1\,J101841.7$-$034131  &      1.7$\pm$0.4&      -&      1.1$\pm$0.3  & 00035818  & 2007 Dec & 2.03 & 0.014$\pm$0.003 \\
XMMSL1\,J103335.6$-$321047  &      1.5$\pm$0.6&      -&      -  & 00035830  & 2006 Nov & 2.26 & $\le$2.57$\times$10$^{-3}$ \\
XMMSL1\,J114354.8$-$690505  &      1.9$\pm$0.4&      1.1$\pm$0.4&     0.8$\pm$0.3 & 00037858  & 2009 Feb & 2.34 & 0.076$\pm$0.006 \\
  XMMSL1\,J115034.4+430453  &      -&      1.5$\pm$0.5&      -    & 00035848  & 2006 Oct & 0.59 & $\le$1.26$\times$10$^{-2}$ \\
XMMSL1\,J120118.8$-$523000  &      2.0$\pm$0.7&      -&      1.5$\pm$0.5  & 00035798  & 2006 Sep, Dec & 2.50 & $\le$2.33$\times$10$^{-3}$ \\
  XMMSL1\,J121730.8+102253  &      2.1$\pm$0.7&      -&      1.6$\pm$0.5    &   00035796  & 2007 Jun & 2.68 & $\le$2.17$\times$10$^{-3}$ \\
  XMMSL1\,J123316.9+213224  &      1.1$\pm$0.5&      1.3$\pm$0.5&      -    & 00035851  & 2007 Mar & 1.62 & $\le$4.30$\times$10$^{-3}$ \\
 XMMSL1\,J125522.0$-$221035 &      1.7$\pm$0.4&      -&      1.1$\pm$0.3   & 00035841  & 2007 Mar, Dec & 2.92 & 0.033$\pm$0.004 \\
                            &     0.6$\pm$0.2&      -&     0.5$\pm$0.2  & & & & \\
XMMSL1\,J131651.2$-$084915  &      1.5$\pm$0.5&      -&      1.2$\pm$0.4   & 00035827  & 2006 Dec & 3.29 & 0.017$\pm$0.002 \\
XMMSL1\,J132442.4$-$712852  &      1.3$\pm$0.4&      -&     0.8$\pm$0.3  & 00037875  & 2009 Feb & 1.85 & $\le$5.04$\times$10$^{-3}$ \\
  XMMSL1\,J134637.9+650319  &      2.0$\pm$0.6&      -&      1.1$\pm$0.4    & 00037855  & 2008 May & 2.12 & $\le$2.74$\times$10$^{-3}$ \\
XMMSL1\,J140743.6$-$430516  &      4.5$\pm$0.7&      1.6$\pm$0.4&      2.7$\pm$0.4 & 00037846  & 2008 Aug & 3.31 & $\le$2.11$\times$10$^{-3}$ \\
  XMMSL1\,J140924.4+675831  &      1.9$\pm$0.7&      -&      1.5$\pm$0.5    & 00035803  & 2006 Oct & 2.84 & $\le$2.28$\times$10$^{-3}$ \\
XMMSL1\,J141843.5$-$293749  &      1.9$\pm$0.6     & -&      1.1$\pm$0.4  & 00035804  & 2006 Dec & 5.78 & 0.042$\pm$0.003 \\
 XMMSL1\,J142022.5$-$384430 &      5.0$\pm$0.7&      -&      4.3$\pm$0.6   & 00037845 & 2009 Dec & 2.71 & $\le$3.12$\times$10$^{-3}$ \\
XMMSL1\,J143651.4$-$090050  &      1.1$\pm$0.3&      -&      3.8$\pm$1.4  & 00035832  & 2006 Sep & 1.34 & 0.045$\pm$0.006 \\
 \end{tabular} 
\end{center}
\end{table*}

\begin{table*}
\contcaption{}
\begin{center}
\begin{tabular}{l c c c l l l l}
XMMSL1\,J145037.6$-$281424  &      2.0$\pm$0.5&      -&      1.8$\pm$0.4  & 00037854  & 2008 Sep & 2.43 & $\le$4.42$\times$10$^{-3}$ \\
   XMMSL1\,J152543.6+672822 &      1.7$\pm$0.6&      1.1$\pm$0.6&      - & 00035812  & 2007 Jan & 2.26 & $\le$4.83$\times$10$^{-3}$ \\ 
 XMMSL1\,J160336.7+774328   &      1.2$\pm$0.4&      1.3$\pm$0.4&      - & 00037879  & 2008 Jun & 3.36 & $\le$2.08$\times$10$^{-3}$ \\
 XMMSL1\,J161944.0+765545   &      2.4$\pm$0.6&      -&     1.0$\pm$0.4 & 00035792  & 2006 Oct & 4.60 & 0.035$\pm$0.003 \\
XMMSL1\,J162136.0+093304    &      2.0$\pm$0.6&      -&      1.1$\pm$0.4 & 00035800  & 2006 Aug, Oct & 7.09 & $\le$8.20$\times$10$^{-4}$ \\
  XMMSL1\,J162533.2+632411  &      1.4$\pm$0.5&      1.2$\pm$0.5&      - & 00035854  & 2007 Jan & 2.68 & 0.007$\pm$0.002 \\
  XMMSL1\,J163439.3+184545  &      1.5$\pm$0.5&      1.5$\pm$0.5&      - & 00037865  & 2008 May & 1.61 & $\le$3.62$\times$10$^{-3}$ \\
  XMMSL1\,J164212.2$-$293051&      1.1$\pm$0.4&      -&      1.0$\pm$0.3 & 00035843  & 2007 Jan & 2.51 & 0.026$\pm$0.004 \\
  XMMSL1\,J164456.7$-$450015&      1.3$\pm$0.3&      1.2$\pm$0.3&      - & 00037876  & 2009 Feb & 3.45 & 2.26$\times$10$^{-3}$ \\
  XMMSL1\,J164859.4+800507  &      1.6$\pm$0.5&      -&     0.7$\pm$0.3 & 00035823  & 2007 Jan & 2.32 & 0.031$\pm$0.004 \\
  XMMSL1\,J170014.4$-$730348&      1.8$\pm$0.6&      -&      1.0$\pm$0.5 & 00035806  & 2006 Sep, 2007 Jun & 5.39 & $\le$1.63$\times$10$^{-3}$ \\
  XMMSL1\,J172700.3+181422  &      2.2$\pm$0.4&     0.7$\pm$0.2&      1.2$\pm$0.2 & 00037852  & 2008 Jul & 2.24 & $\le$2.59$\times$10$^{-3}$ \\
  XMMSL1\,J173637.8$-$193611&      2.0$\pm$0.3&      -&      1.3$\pm$0.3 & 00035799  & 2007 Mar, 2009 Feb & 4.98 & $\le$1.92$\times$10$^{-3}$\\
  XMMSL1\,J175542.2+624903  &     0.9$\pm$0.5&      -&     0.7$\pm$0.3 & 00035844  & 2007 Feb, Mar & 2.72 & 0.039$\pm$0.004 \\
                            &      1.2$\pm$0.4&      -&      1.0$\pm$0.3 &            & & & \\
  XMMSL1\,J181659.8$-$254219&      1.6$\pm$0.5&      1.6$\pm$0.5&      - & 00035819  & 2008 Feb & 0.67 & $\le$8.64$\times$10$^{-3}$ \\
  XMMSL1\,J182707.5$-$465626&      2.2$\pm$0.6&      -&      1.2$\pm$0.5 & 00035794  & 2007 Jun, 2008 Jun & 3.41 & 0.007$\pm$0.002 \\
  XMMSL1\,J182933.3+175619  &      1.3$\pm$0.4&      -&     0.7$\pm$0.3 & 00037880  & 2008 Dec, 2009 May & 3.53 & $\le$2.10$\times$10$^{-3}$ \\
  XMMSL1\,J183233.0$-$112539&      1.0$\pm$0.3&      1.3$\pm$0.4&      - & 00035850  & 2006 Oct& 2.26 & $\le$5.02$\times$10$^{-3}$ \\
  XMMSL1\,J183642.8$-$583857&      -&      1.0$\pm$0.4&      - & 00035859  & 2007 Mar & 3.68 & $\le$1.89$\times$10$^{-3}$  \\
  XMMSL1\,J185314.2$-$363057&      2.9$\pm$0.7&      -&      2.2$\pm$0.4 & 00035788  & 2006 Oct & 2.71 & 0.044$\pm$0.004 \\
  XMMSL1\,J185608.5$-$430320&      1.4$\pm$0.6&      -&      1.2$\pm$0.5 & 00035839  & 2007 Apr, Jul & 4.22 & $\le$1.65$\times$10$^{-3}$ \\
  XMMSL1\,J191028.2+495606  &      1.6$\pm$0.3&     0.5$\pm$0.2&      1.1$\pm$0.2 & 00035825  & 2006 Nov & 2.16 &$\le$4.06$\times$10$^{-3}$ \\
  XMMSL1\,J200203.1$-$055152&      4.4$\pm$1.3&      -&      2.7$\pm$0.9 & 00035784  & 2006 Oct & 2.18 & $\le$6.23$\times$10$^{-3}$ \\
  XMMSL1\,J203044.2$-$484718&      2.4$\pm$1.1&      -&      - & 00035790  & 2006 Nov, 2007 Mar & 5.44 & $\le$1.71$\times$10$^{-3}$ \\
  XMMSL1\,J204033.2+482749  &      1.7$\pm$0.5&     0.9$\pm$0.4&      - & 00035816  & 2007 Mar & 2.49 & $\le$3.27$\times$10$^{-3}$\\
  XMMSL1\,J204142.3+185258  &      2.8$\pm$0.6&      1.5$\pm$0.5&      1.1$\pm$0.4 & 00037848  & 2008 Jun & 2.05 & $\le$2.83$\times$10$^{-3}$\\
  XMMSL1\,J205542.2$-$115756&      1.5$\pm$0.5&      -&      1.2$\pm$0.4 & 00037867  & 2008 Jun & 3.60 & 0.020$\pm$0.003 \\
  XMMSL1\,J211420.7+252419  &      5.6$\pm$0.7&     0.7$\pm$0.2&      4.1$\pm$0.5 & 00035783  & 2006 Dec & 2.14 & 0.110$\pm$0.008\\
  XMMSL1\,J211506.4+305811  &      1.7$\pm$0.6&      -&      - & 00035817  & 2007 Mar, Apr & 3.28 & $\le$1.77$\times$10$^{-3}$ \\
  XMMSL1\,J213537.3+024834  &      1.9$\pm$0.5&      1.6$\pm$0.5&      - & 00037859  & 2009 Apr& 2.24 & $\le$2.60$\times$10$^{-3}$\\
  XMMSL1\,J215303.1$-$173633&      1.6$\pm$0.6&      -&      - & 00035822  & 2007 Apr& 2.74 & $\le$2.54$\times$10$^{-3}$ \\
  XMMSL1\,J215905.6$-$201604&      2.1$\pm$0.4&      -&      1.8$\pm$0.3 & 00037853  & 2008 Dec & 3.81 & 0.006$\pm$0.001 \\
  XMMSL1\,J230652.9+213159  &      1.6$\pm$0.7&      -&      - & 00035820  & 2007 May & 2.90 & $\le$2.01$\times$10$^{-3}$ \\
  XMMSL1\,J230937.0$-$522529&     0.9$\pm$0.3&      1.1$\pm$0.3&      - & 00035857  & 2006 Dec & 4.13 & $\le$1.57$\times$10$^{-3}$ \\
  XMMSL1\,J235604.5+400726  &      1.5$\pm$0.4&      -&     0.9$\pm$0.3 & 00037864  & 2008 Jul & 2.31 & $\le$2.52$\times$10$^{-3}$ \\
\end{tabular} 
\end{center}
\end{table*}

\section{X-ray results}
We analysed the data according to the recipes given in \cite{Evans} and based on the publicly available {\it Swift} data analysis tools at \url{http://www.swift.ac.uk/user_objects}. All observations were reduced and analysed homogeneously, using the {\it Swift} software version 3.4 (HEASOFT 6.7) and latest calibration files as of 2009 November 1. For a detection we require a source significance above the background of 3$\sigma$.
To be considered a match, the XRT-detected source position must agree with the Slew Survey position when adopting 3$\sigma$ positional uncertainties. We detect 27 of the 94 observed sources with the XRT, corresponding to a detection rate of 29\%. The mean count rate for the detected sources is 0.038 count s$^{-1}$, while the mean detection limit is 0.003 count s$^{-1}$ (Table \ref{tab:obs}), corresponding to a 0.3--10\,keV flux of 10$^{-13}$ erg cm$^{-2}$ s$^{-1}$ for an absorbed power law with $\Gamma = 2$ and $N_{\rm H} = 10^{21}$ cm$^{-2}$.

\subsection{Position improvement} \label{positions}
\begin{figure}
\begin{center}
\includegraphics[width=6cm, angle=90]{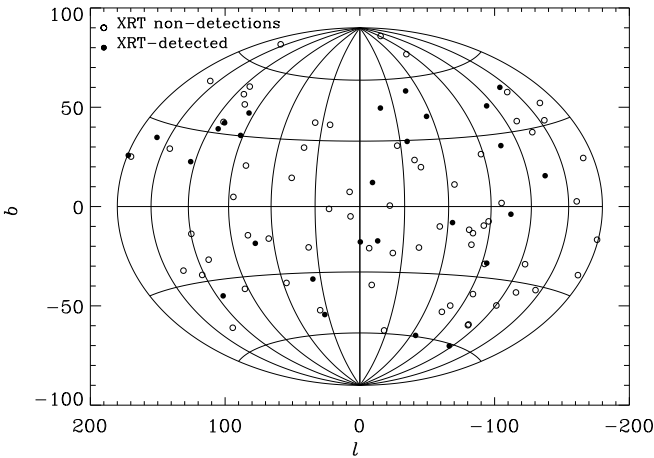}
\caption{Aitoff projection in Galactic coordinates of the distribution of unidentified {\it XMM-Newton} Slew Survey sources in our sample, both {\it Swift} XRT-detected (using the new XRT positions) and undetected (using the {\it XMM-Newton} Slew Survey positions).}
\label{aitoff}
\end{center}
\end{figure}

From the {\it XMM-Newton} Slew Survey, X-ray positions were measured for this sample to accuracies of $\sim$10$''$ to $\sim$2$'$ (radius, 1$\sigma$, including systematic error). For the detected sources the mean 90\% confidence X-ray positional error radius derived from the {\it Swift} data is 2.9$''$ (statistical+systematic, Table \ref{tab:positions}), i.e. significantly improved compared to the 18.9$''$ mean {\it XMM-Newton} Slew Survey uncertainty. Positions determined by the XRT can be improved in both accuracy and precision by using the UVOT to accurately determine the spacecraft pointing \citep[see][for full details of this procedure]{Goad,Evans}; this method was used where possible. All but four sources have UVOT-enhanced XRT positions. The XRT positions are centred between 0.8$''$ and 34$''$ from the {\it XMM-Newton} Slew Survey central positions. The detected sources are not concentrated along the Galactic Plane nor at the Galactic Centre, but appear to be randomly distributed in the sky, and are not distributed differently to the undetected sources (Fig. \ref{aitoff}). 

\begin{table*}
\caption{Source positions and associated errors (radius, 90\% containment) derived from the XRT observations. All XRT positions have been UVOT-enhanced, except where indicated with $^*$.}
\label{tab:positions}
\begin{center}
\begin{tabular}{l c l c l l}
Source & {\it XMM} position & error & XRT position & error & position difference \\
XMMSL1\,J& (deg) & ($''$) & (deg) & ($''$) & ($''$) \\\hline
002202.9$+$254004 &  5.51246, 25.66780 & 11   &5.51330, 25.66779 & 1.5 & 2.7 \\
010654.8$+$802740 &  16.72840, 80.46115 &10   &16.72612, 80.45951& 3.9& 6.1 \\
012240.2$-$570859 &   20.66791, -57.14976& 19 &20.67286, -57.15126&3.6&11.1 \\
030006.6$-$381617 &   45.02795, -38.27129& 10 &45.02783, -38.27109& 2.1 & 0.8 \\
064109.2$-$565542 &  100.28859, -56.92840& 20 &100.28459, -56.93256& 2.3& 16.9 \\
065525.2$+$370815 &  103.85532, 37.13737& 10  &103.85670, 37.13786&2.9 & 4.3 \\
070846.2$+$554905 &   107.19079, 55.81748& 12 &107.19302, 55.81767& 1.6& 4.6 \\
075818.9$-$062723  & 119.57912, -6.45650& 28  &119.57775, -6.45644& 3.5& 4.9 \\
080849.0$-$383803 &    122.20423, -38.63413&16&122.19943, -38.63185&1.7&15.8 \\
094551.3$-$194352 &    146.46383, -19.73120&17&146.46285, -19.73380& 1.5& 9.9\\
095336.4$+$161231& 148.40175, 16.20851& 11    &148.40366, 16.20801&2.6&6.8\\
101841.7$-$034131&    154.67358, -3.69182& 11 &154.67482, -3.69063& 2.7& 6.2\\       
114354.8$-$690505&   175.97849, -69.08476& 15 &175.97225, -69.08709&3.9$^*$&11.6 \\
125522.0$-$221035&    193.84203, -22.17633& 11&193.84293, -22.17791& 3.7$^*$&6.4 \\
131651.2$-$084915&   199.21291, -8.82081 &11  &199.21360, -8.82178& 2.1& 4.3 \\
141843.5$-$293749&     214.68130, -29.63044&12&214.68451, -29.63055& 1.5& 10.1\\
143651.4$-$090050&    219.21416, -9.01379& 26 &219.22130, -9.00926 & 3.7& 30.2 \\
161944.0$+$765545&  244.93372, 76.92922& 22   &244.91138, 76.92113& 1.8& 34.3\\
162533.2$+$632411 & 246.38844, 63.40311& 20   &246.38694, 63.39842& 4.0& 17.1\\
164212.2$-$293051 &   250.55056, -29.51433& 11&250.55137, -29.51379& 2.0& 3.2\\
164859.4$+$800507& 252.24682, 80.08518& 26   &252.18727, 80.08678 & 9.6 & 37.4 \\
175542.2$+$624903&268.93855, 62.82350&12      &268.94182, 62.82487 & 1.7&7.3 \\
182707.5$-$465626 &   276.78128, -46.94063& 19&276.78290, -46.94032&4.3$^*$& 4.1 \\
185314.2$-$363057 &   283.30962, -36.51566& 12&283.30667, -36.51574& 2.2& 8.5 \\
205542.2$-$115756 &  313.92622, -11.96551& 11 &313.92731, -11.96643 & 2.0 & 5.1 \\
211420.7$+$252419 &  318.58667, 25.40539& 118  &318.58790, 25.40603 & 1.8 & 4.6 \\
215905.6$-$201604 & 329.77282, -20.26792& 11  &329.77492, -20.26781& 4.1$^*$ & 7.1 \\
\end{tabular} 
\end{center}
\end{table*}

\subsection{Comparison with previous high energy catalogues} \label{rosatsec}
Using the newly derived {\it Swift} positions for the XRT-detected sources, we searched all major high energy catalogues, including data from {\it Einstein}, {\it Ginga}, {\it ROSAT}, {\it ASCA}, {\it XMM-Newton}, {\it Chandra}, {\it BATSE}, {\it GRANAT}, {\it SAS-2} and {\it EUVE}. This results in ten sources with one or more possible {\it ROSAT} counterparts (Table \ref{tab:rosat}). In addition, the revised position of one source suggests association with another {\it XMM-Newton} Slew Survey object (see Table \ref{tab:rosat}). We have used the 3$\sigma$ error radii on the X-ray positions to search for a match, which includes the 6$''$ (1$\sigma$) systematic error in the case of RASS, while we note that the 1.4$''$ (90\%) systematic error for enhanced XRT positions (3.5$''$ where enhancement was not possible) is already included in all reported XRT positions. One of the {\it ROSAT} matches has been classified as a Type I AGN, from both X-ray and optical observations, and lies at a redshift of $z = 0.236$ \citep{Gioia}. Another {\it ROSAT} match has been classified as an F-G type star in the Hamburg/RASS Catalogue of optical identifications V3.0 \citep{Zickgraf}. The XRT-detected source coincident with this RASS star is also coincident with a second Slew Survey source not included in our sample: only one X-ray source is found in the $\sim$17$\times$17 arcmin XRT field of view, hence these two Slew Survey sources and one RASS source may all be one and the same. In order to verify this, we include this XRT-detected source in all further analysis presented here. No other {\it ROSAT} matches have existing classifications. We note that one of our sample sources, XMMSL1\,J162533.2$+$632411, was not detected in the soft band in the Slew Survey and is spectrally the hardest among the XRT detected sample, making the {\it ROSAT} association uncertain. 
\begin{table*}
\caption{Possible {\it ROSAT} associations, displaying agreement within the 3$\sigma$ positional uncertainties.}
\label{tab:rosat}
\begin{center}
\begin{tabular}{ l l l l l l}
{\it XMM} Slew name  &  {\it ROSAT} name & {\it ROSAT} pos. err.& XRT pos. err.& pos. diff. & ID (ref.) / Comments\\ 
           XMMSL1\,J                                & &($''$)  &($''$) &($''$)&\\ \hline 
012240.2$-$570859 & 1RXS\,J012245.0$-$570901 & 75& 6.6& 28.8 &\\
064109.2$-$565542&1RXS\,J064106.5$-$565610 & 39 & 4.2& 19.5 & \\
114354.8$-$690505&1RXH\,J114353.5$-$690513 & 30 &7.1&1.1&\\ 
&1RXH\,J114353.3$-$690506 & 24 & &7.5  &\\
&1RXH\,J114351.8$-$690505& 18 &  & 10.8&\\
 141843.5$-$293749 & 1RXS\,J141846.1$-$293748 & 33 & 2.7& 23.8 & \\
 143651.4$-$090050& 1RXS\,J143653.7$-$090004 & 111 & 6.7& 30.1 & \\
 161944.0$+$765545& 1RXS\,J161939.9$+$765515 & 24 & 3.3& 4.0 & Star F-G \citep{Zickgraf}\\
                  &                       &    &    &     & associated with \\
                  &                       &    &    &     & XMMSL1\,J161935.7$+$765508\\
 162533.2$+$632411& 1RXS\,J162535.1$+$632333 & 63 & 7.3& 26.1 & No soft band detection in Slew\\
 164212.2$-$293051& 1RXS\,J164216.5$-$293035 & 129 & 3.6& 56.4 & \\
 164859.4$+$800507& 1RXS\,J164843.5$+$800506 & 27 & 17.5& 7.4  &\\
 175542.2$+$624903& 1RXS\,J175546.2$+$624927 & 21 & 3.1& 2.8  &AGN Type I, $z = 0.236$ \\
                  &                       &    &    &     & \citep{Gioia}\\ \hline

{\it XMM} Slew name  &  {\it XMM} Slew name & {\it XMM} pos. err.& XRT pos. err.& pos. diff. & ID (ref.) / Comments\\ 
       XMMSL1\,J (this sample)     &   (possible counterpart)   &(counterpart, $''$)  &($''$) &($''$)&\\ \hline 
 161944.0$+$765545& XMMSL1\,J161935.7$+$765508 & 32 & 3.3 & 13& Star F-G \citep{Zickgraf}\\
                  &                       &    &    &     & associated with \\
                  &                       &    &    &     &  1RXS\,J161939.9$+$765515\\ \hline
\end{tabular}
\end{center}
\end{table*}

\subsection{X-ray spectral analysis}
We created one XRT spectrum per detected source, fitted using the C-statistic \citep{Cash} in {\sc Xspec}.
The X-ray spectral model employed was an absorbed power law, where the single absorption component $N_{\rm H}$ at $z \equiv 0$ was left to vary freely rather than set to the Galactic value to allow for sources within the Galaxy as well as extragalactic objects. Results of the spectral fits are given in Table \ref{tab:spec}, where we also list the expected Galactic extinction from the Leiden Argentine Bonn HI maps \citep[LAB,][]{Kalberla}. In one case there was no acceptable fit.

\begin{figure}
\begin{center}
\includegraphics[width=6cm, angle=90]{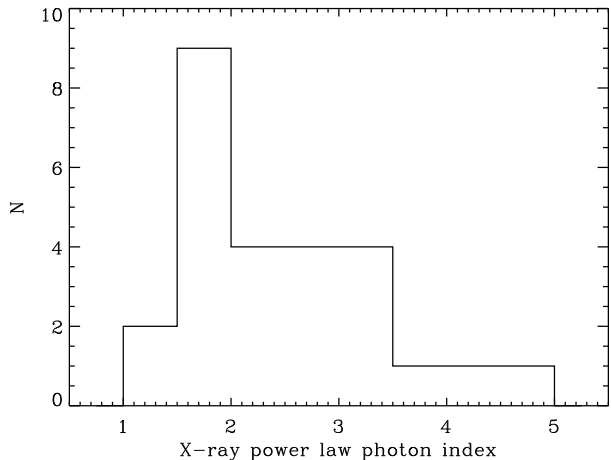}
\caption{Distribution of XRT best-fitting power law photon indices for the detected sources, excepting XMMSL1\,J010654.8$+$802740.}
\label{gammas}
\end{center}
\end{figure}

The power law photon indices cluster around $\Gamma= 1.5-2.0$ (Fig. \ref{gammas}), typical of Active Galactic Nuclei \citep[AGN,][]{Mateos05,Mainieri,Mateos2010}. The measured equivalent hydrogen column densities with this model are consistent with or in excess of the Galactic value for all sources except XMMSL1\,J080849.0$-$383803: this source has a column far lower than the expected Galactic value and therefore probably lies close-by, within our Galaxy. We discuss the accuracy of the Galactic column in this direction in Appendix A9, where we conclude that the source is Galactic.  

A power law fit to one of the sample sources, XMMSL1\,J161944.0$+$765545, results in a very soft photon index, $\Gamma=4.58^{+0.86}_{-0.62}$, and no counts are detected at $>$ 5\,keV. 
In Table \ref{tab:rosat} we have shown that XMMSL1\,J161944.0$+$765545 is associated with the {\it ROSAT}-discovered source 1RXS\,J161939.9$+$765515, classed as an F-G star. An absorbed {\sc MEKAL} model fit to the XRT spectrum results in a plasma temperature of $kT = 1.2^{+0.1}_{-0.2}$\,keV and a total absorbing column of $\le$~2$\times$10$^{20}$ cm$^{-2}$ with C-statistic (dof) = 93 (90). The measured upper limit on the total column density is lower than the mean Galactic column of 4.1$\times$10$^{20}$ cm$^{-2}$ in that direction, consistent with its classification. 

\begin{table*}
\caption{ Results of absorbed power law fits to the X-ray spectra of XRT detected sources. $N_{\rm H,Gal}$ is a weighted average at the XRT position \citep{Kalberla}. Fluxes are given for the interval 0.3--10 keV. We also list the total number of source counts in the final column. \newline$^{*}$ flux is taken from an unconstrained absorbed power law fit and so is approximate. This fit had a power law photon index of $\Gamma \sim 9.3$ and absorption $N_{\rm H} \sim 10^{22}$ cm$^{-2}$; blackbody and mekal fits also resulted in no acceptable solutions.}
\label{tab:spec}
\begin{center}
\begin{tabular}{l l l l l l l l l }
Source & $\Gamma$ & $N_{\rm H}$ & $N_{\rm H,Gal}$& observed flux & unabsorbed flux& C-stat (dof)& N$_{\rm counts}$ \\
XMMSL1\,J& &$\times 10^{20}$&$\times 10^{20}$&$\times 10^{-12}$&$\times 10^{-12}$ & &  \\
&& (cm$^{-2}$) &  (cm$^{-2}$)& (erg cm$^{-2}$ s$^{-1}$) & (erg cm$^{-2}$ s$^{-1}$) & &\\
 \hline
002202.9$+$254004 &  1.69$^{+0.14}_{-0.19}$& 1.6$^{+4.3}_{-1.6}$&3.03 & 5.9$^{+1.1}_{-1.5}$& 6.1& 209.24 (191)& 328\\
010654.8$+$802740 &  \multicolumn{2}{l}{no acceptable fit}&16.0&0.34$^{*}$&-&-&43 \\
012240.2$-$570859 &  3.74$^{+0.83}_{-0.74}$& 10.0$^{+10.6}_{-8.1}$ &2.96 & 1.21$^{+0.38}_{-0.71}$& 3.1& 55.26 (70)& 123\\
030006.6$-$381617 &  1.91$^{+0.61}_{-0.33}$& unconstrained&1.85 & 0.83$^{+0.34}_{-0.83}$& 0.83& 25.76 (33)&42\\
064109.2$-$565542 &  1.73$^{+0.27}_{-0.35}$& 8.2$^{+5.8}_{-7.8}$&5.81 & 1.62$^{+0.45}_{-0.84}$& 1.9&104.27 (122)& 216\\
065525.2$+$370815 & 2.5$^{+2.4}_{-1.3}$& 22$^{+51}_{-22}$&11.3 & 0.24$^{+0.15}_{-0.24}$& 0.48& 19.95 (18)& 87\\
070846.2$+$554905 & 1.80$^{+0.25}_{-0.26}$& 8.2$^{+3.3}_{-6.0}$&5.06&2.97$^{+0.59}_{-1.22}$& 3.5&147.11 (165)&257\\
075818.9$-$062723 & 1.63$^{+0.45}_{-0.42}$& 9.4$^{+11.7}_{-9.4}$&8.21& 0.97$^{+0.33}_{-0.87}$&1.1&73.13 (81)& 107\\
080849.0$-$383803 & 3.24$^{+0.74}_{-0.59}$& 14$^{+11}_{-8}$&75.8&1.76$^{+0.54}_{-1.25}$& 4.4&69.22 (75)& 112\\
094551.3$-$194352 & 1.66$^{+0.19}_{-0.21}$& 24$^{+3}_{-7}$&3.83&2.64$^{+0.45}_{-0.70}$&3.4&266.09 (273)& 500\\
095336.4$+$161231 & 2.31$^{+1.04}_{-0.72}$& 8.7$^{+21.0}_{-8.7}$&3.34&0.61$^{+0.32}_{-0.61}$& 0.84&31.26 (32)&41\\
101841.7$-$034131 & 2.8$^{+2.5}_{-1.5}$& 35$^{+56}_{-33}$&3.81&0.49$^{+0.31}_{-0.49}$&1.6&22.33 (22)& 28\\
114354.8$-$690505 & 1.66$^{+0.30}_{-0.30}$& 29$^{+12}_{-11}$ &18.0&5.8$^{+1.3}_{-2.2}$& 7.8&144.64 (180)&177\\
125522.0$-$221035 & 1.44$^{+0.41}_{-0.39}$& 5.2$^{+12.7}_{-5.2}$&6.47&1.80$^{+0.70}_{-1.45}$&1.9&84.69 (57)&96\\
131651.2$-$084915 & 2.89$^{+1.25}_{-0.64}$ & 3.5$^{+15.3}_{-3.2}$&2.37& 0.48$^{+0.24}_{-0.48}$& 0.62&46.40 (39)&56\\
141843.5$-$293749 & 1.86$^{+0.27}_{-0.28}$& 8.4$^{+5.8}_{-6.3}$& 4.95&1.76$^{+0.42}_{-0.67}$& 2.1& 100.84 (154)&  242\\
143651.4$-$090050 & 3.1$^{+1.6}_{-1.4}$& 31$^{+32}_{-16}$&5.80&1.37$^{+0.54}_{-1.37}$& 5.1& 59.09 (48)& 60\\
161944.0$+$765545 & 4.58$^{+0.86}_{-0.62}$ & 41$^{+14}_{-16}$&4.10& 0.83$^{+0.24}_{-0.78}$& 14.0&80.23 (90)&161\\
 162533.2$+$632411& 1.05$^{+0.82}_{-0.63}$&unconstrained&1.86&0.54$^{+0.47}_{-0.54}$&0.54&14.89 (13)&  18\\
164212.2$-$293051 & 4.2$^{+1.3}_{-1.1}$& 29$^{+20}_{-19}$&13.5&0.65$^{+0.28}_{-0.65}$ &4.9&45.85 (49)& 65\\
164859.4$+$800507 & 2.47$^{+0.69}_{-0.60}$& 17$^{+17}_{-13}$&4.52&1.07$^{+0.41}_{-1.07}$&1.9&52.10 (56)& 72\\
 175542.2$+$624903& 2.10$^{+0.39}_{-0.23}$& unconstrained &3.25&1.42$^{+0.38}_{-0.81}$& 1.4& 64.36 (72)& 106\\
182707.5$-$465626 & 1.83$^{+0.92}_{-0.48}$&unconstrained&5.99&0.29$^{+0.16}_{-0.29}$&0.29&16.56 (21)&  23\\
185314.2$-$363057 &2.78$^{+0.76}_{-0.68}$&16$^{+15}_{-6}$&6.65&1.28$^{+0.42}_{-1.13}$& 2.7&  75.60 (80)&119\\
205542.2$-$115756 & 3.09$^{+1.21}_{-0.69}$&19$^{+18}_{-13}$&4.49&0.57$^{+0.23}_{-0.57}$&1.6&37.15 (52)&  72\\
211420.7$+$252419 & 2.60$^{+0.25}_{-0.33}$&11$^{+7}_{-6}$&7.53&  3.39$^{+0.70}_{-0.95}$& 5.7&83.51 (146)&235\\
215905.6$-$201604 & 3.2$^{+3.2}_{-1.4}$& 17$^{+47}_{-17}$&2.47&0.17$^{+0.12}_{-0.17}$&0.49&13.57 (18)&22\\
\end{tabular} 
\end{center}
\end{table*}

\subsection{X-ray variability}\label{sec:variability}
\begin{figure*}
\begin{center}
\includegraphics[width=6cm, angle=90]{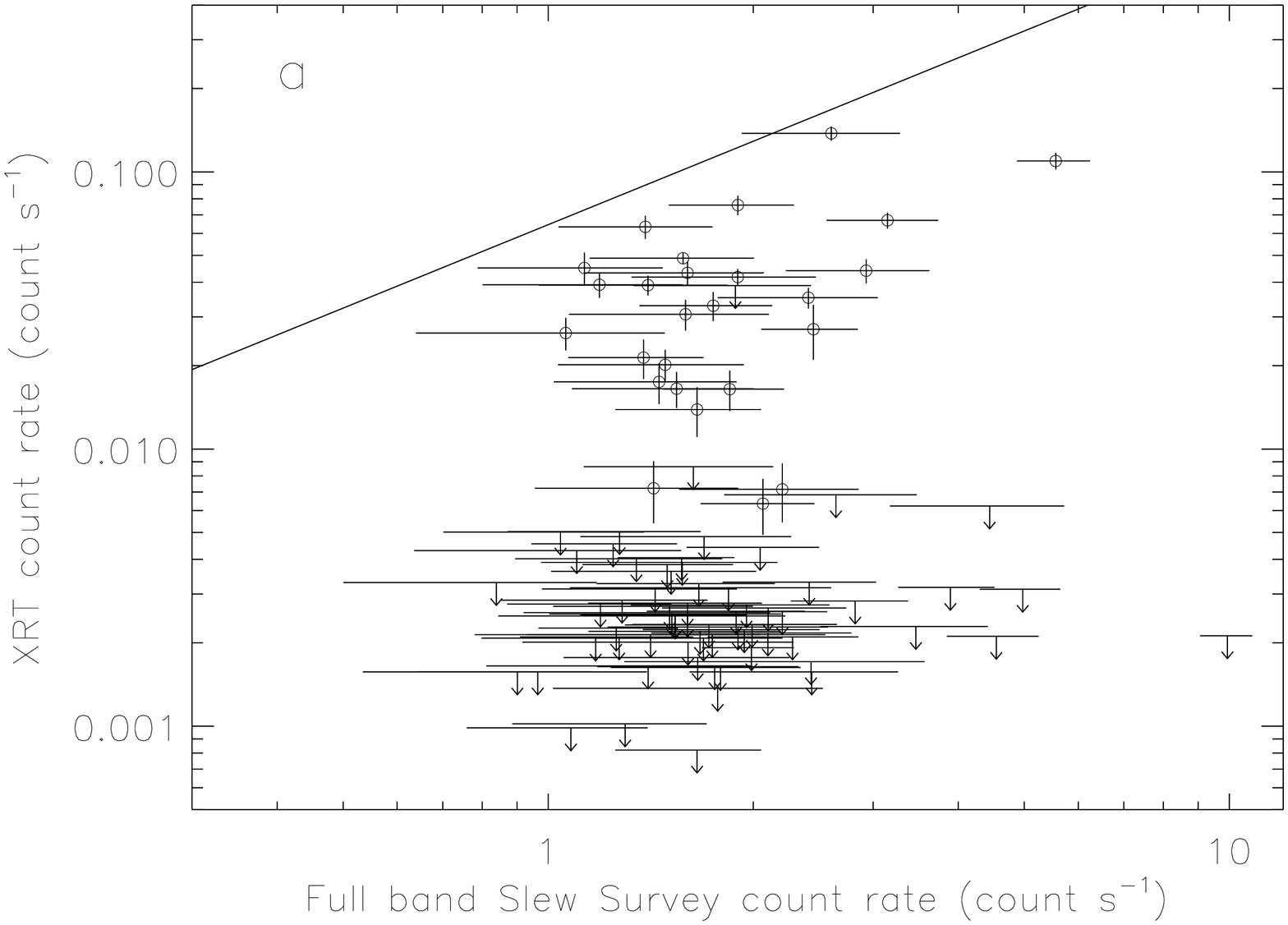}
\hspace{0.3cm}
\includegraphics[width=6cm, angle=90]{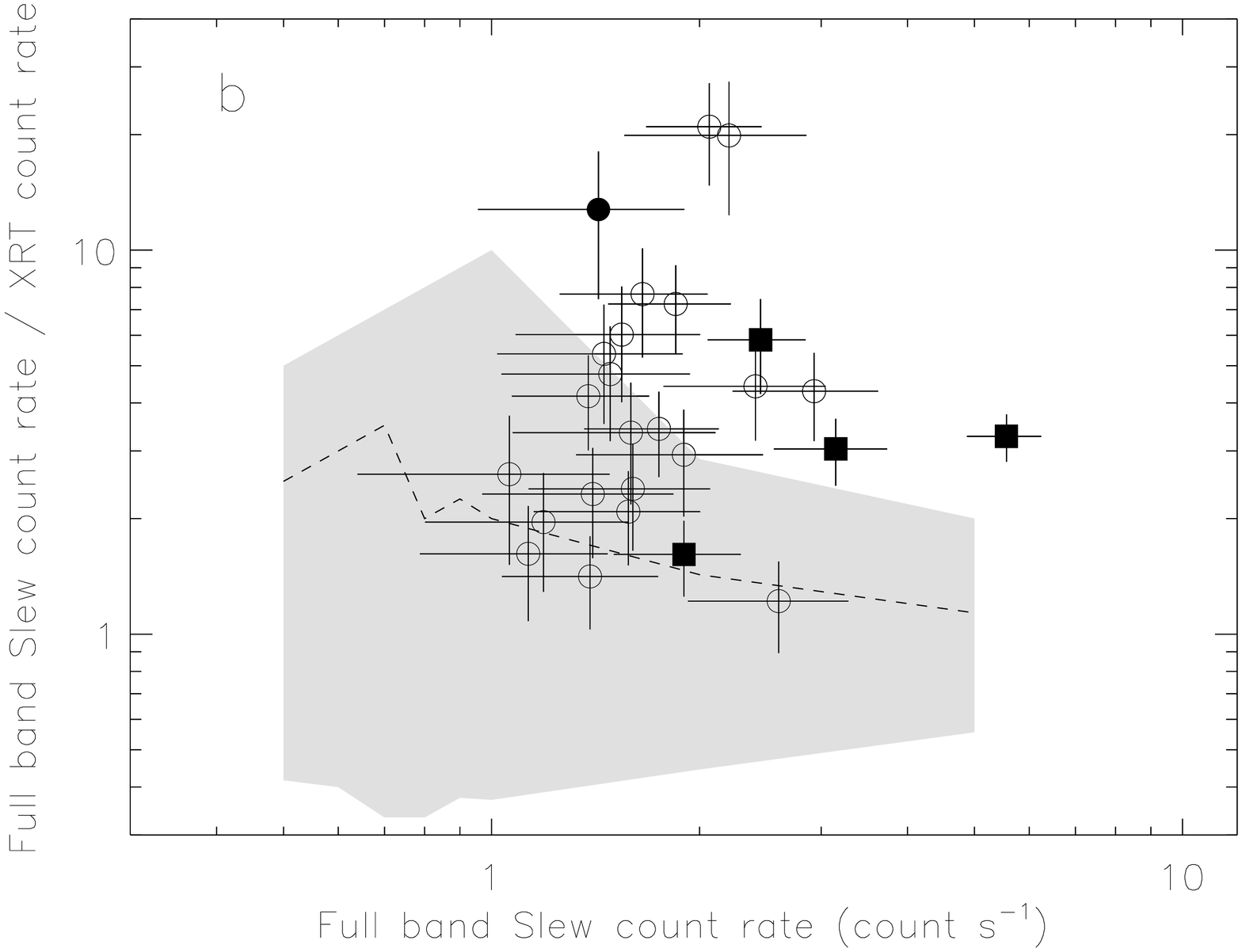}
\caption{a) Observed {\it XMM-Newton} pn count rate versus observed {\it Swift} XRT count rate (open circles and 1$\sigma$ error bars for XRT detections, upper limits for XRT non-detections). The solid line shows a 15.5:1 ratio which is expected for the transformation between count rate for the two instruments (Section \ref{sec:variability}).\newline b) The ratio between the {\it XMM-Newton} Slew Survey and XRT full band count rates after correcting for the expected 15.5:1 count rate ratio factor. Open circles denote soft band Slew Survey detections, filled circles denote hard band Slew Survey detections and filled squares denote hard+soft band Slew Survey detections. The shaded area and dashed line indicate the values expected (full range and at peak, respectively, see Section \ref{sec:variability} for details) for non-variable sources from Eddington Bias simulations performed for the {\it XMM-Newton} Slew hard band survey. The location of points both within and outside of the shaded area shows that the XRT-detected population likely comprises both steady and substantially variable sources.}
\label{countrates}
\end{center}
\end{figure*}

The {\it Swift} 0.3--10\,keV XRT count rates and upper limits all lie below those expected, given the catalogued full-band count rates in the {\it XMM-Newton} Slew Survey performed 1.6--5.4 years earlier (Fig. \ref{countrates}a). 
Many of the {\it Swift} XRT
observations to date provide a null detection suggesting that the
source has dropped in X-ray flux by a factor of 10 to 100 or
more.

In order to compare the {\it Swift} XRT and {\it XMM-Newton} observations, we need to understand the relationship between the Slew Survey count rates and XRT count rates. {\sc Xspec} simulations show that the relative count rate ratio expected
between {\it XMM-Newton} EPIC pn and XRT PC mode is 15.5:1 for a typical AGN spectrum with an X-ray absorbing column of $N_{\rm H}=3\times 10^{20}$ cm$^{-2}$ and a power law photon index of $\Gamma=1.7$. This may be uncertain by $\pm$10\% considering instrument cross-calibration. After accounting for this factor we still find that all sources but one have lower than expected XRT count rates at least at the 1$\sigma$ level and assuming constant source flux.

In addition, the Eddington Bias \citep{Eddington} should be taken into account. This can boost the true count rates of sources at or near the detection limit. Simulations performed in order to understand the {\it XMM-Newton} hard band survey show that those count rates can be overestimated by a factor of two or more in flux-limited surveys such as this. Warwick et al. (2010, in preparation) have quantified this effect for the {\it XMM-Newton} hard band slew survey (2--12 keV), by using the hard band $\log N - \log S$ to create source distributions, simulating Slew Survey observations of these and measuring the resultant observed count rates. For a given observed count rate they derive the true count rate distribution, which peaks a factor of 2 below the observed count rate for detections with a low
number of counts ($<$8). Using these distributions we find an average conversion from observed to true count rates which could be applied to the Slew Survey count rates (Fig. \ref{countrates}b). We stress that this is at best a first order approximation when applied to the full-band count rates.  

When we compare the observed count rates of the sample in the {\it XMM-Newton} Slew Survey and in the {\it Swift} XRT follow-up, our sample appears to split into two groups (Fig. \ref{countrates}a). One group shows a similar count rate to that expected from the Slew Survey if count rate conversion and biases are taken into account, while the other group is comprised largely of the non-detections: this latter group of sources must, if real, be significantly variable to have been seen in the Slew Survey but not in the {\it Swift} XRT observations. According to the K-S test, the XRT count rates for the detected sources and the upper limits on the XRT count rates for the non-detected sources are significantly different and likely not drawn from the same distribution. We discuss the XRT non-detected population further in Section \ref{sec:undetected}.

Four of the XRT-detected sources show significant X-ray variability on the basis of multiple XRT observations, on time scales of days to months and by factors of 1.4--2. These are XMMSL1\,J002202.9$+$254004, XMMSL1\,J094551.3$-$194352, XMMSL1\,J095336.4$+$161231 and XMMSL1\,J175542.2$+$624903.
A comparison of the average XRT soft band (0.3--2 keV) observed fluxes for XRT detected sources with the RASS fluxes or upper limits identifies 3 sources which have varied by a factor 3 or more on 15--19 year time scales: XMMSL1\,J002202.9$+$254004, XMMSL1\,J030006.6$-$381617 and XMMSL1\,J094551.3$-$194352. The population as a whole includes both increases and decreases in flux since the RASS observations.

\section{Broadband characterisation of X-ray detected sources}
\subsection{Hard X-ray detections with \emph{Swift} BAT} \label{bat}
The wide-field of the BAT hard X-ray detector on-board
{\it Swift} means that our sample
sources were in the BAT field of view on more occasions
than the XRT observations listed in Table \ref{tab:obs}; these data have
been compiled in the BAT 58-month Survey (Baumgartner et al.
2010, in preparation). Although none of the sources were detected above the 4.8$\sigma$ survey
threshold, five sources were detected at $> 3.0 \sigma$ significance. These sources and their 15--150\,keV BAT
count rates are listed in Table \ref{tab:batobservations}. Two of the five sources, XMMSL1\,J002202.9$+$254004 and
XMMSL1\,J185608.5$-$430320, were also detected by the XRT. The limiting BAT count rate on
the non-detected sources is typically 2.4$\times$10$^{−5}$ ct cm$^{-2}$ s$^{-1}$
(1$\sigma$, $\sim$0.1 mCrab). All but one (XMMSL1\,J093738.4$-$654445) of the five sources are at high Galactic latitude ($|b| > 20^{\circ}$ ). While X-ray binaries, pulsars, magnetic cataclysmic variables (CVs), and Be/symbiotic stars can show very high energy X-ray and $\gamma$-ray emission, extragalactic sources are likely to be more numerous among the BAT detections. For example, \cite{Landi} carried out {\it Swift} follow-up observations of 20
unidentified Integral/IBIS sources and found that eleven of
these could be classified as extragalactic - AGN, QSOs and
a LINER. Only one of their sample was confirmed to be a
Galactic object.

The two BAT-detected sources that are also seen
in the X-ray band with XRT have coincident
optical sources which lie at the faint end of the sample range
and have no measured proper motion (Sections \ref{sec:opt1},\ref{sec:opt}). We suggest these are likely extragalactic
jet-dominated sources such as blazars. The low BAT detection rate
we find here is inconsistent with these sources being heavily obscured (Compton thick) AGN.
\cite{Winter} show that the BAT 9-month AGN Survey is complete down to a 2--10\,keV flux of
1.0 $\times$10$^{-11}$ erg cm$^{-2}$ s$^{-1}$ with a 4.8$\sigma$ threshold \citep[Fig. 15 of][]{Winter}. Since the BAT
Survey sensitivity is dominated by statistics, the 58-month survey will go deeper by a factor
of 2.5, and if we accept sources down to 3.0$\sigma$, BAT should detect all AGN down to 2.4$\times$10$^{-12}$ erg cm$^{-2}$ s$^{-1}$, which is below the $\sim$4$\times$10$^{-12}$ 2--12\,keV sensitivity of the {\it XMM-Newton} Slew Survey. Also it is
unlikely that more than a small fraction of the AGN would be so variable as to be bright at the
time of the {\it XMM-Newton} Slew Survey detection and undetectable by either XRT or BAT a few years later. Blazars, however, are
known to have bright X-ray flares and tend to have a soft spectrum in the BAT energy range,
making them relatively difficult to detect.

Four of the five BAT-detected sources were detected only in the soft and full bands in the Slew Survey. One source, XMMSL1\,J093738.4$-$654445, was not detected in the soft band but only in the hard and full bands in the Slew Survey. Hard band only (2--12 keV) Slew Survey detections amount to $\sim$20\% of the full sample, most of which are not detected with {\it Swift} XRT rendering the nature of this population rather difficult to determine (but see Section \ref{sec:undetected}). We note that a hard-only Slew Survey source not a member of our sample, XMMSL1\,J171900.4$-$353217 which has been designated as a hard X-ray flash, was observed and detected with {\it Swift} XRT \citep{Read10,ArmasPadilla}. It is located in the Galactic Ridge and has been shown to be transient in nature through multiple X-ray observations. The hard BAT-detected Slew Survey source in this sample, XMMSL1\,J093738.4$-$654445, lies at low Galactic latitude, $b=-10$, and could be a Galactic hard X-ray transient candidate, while the remaining members could be blazars or other AGN types.
\begin{table}
\caption{BAT detections in the 58-month hard X-ray survey. The significance of the detection is given by $\sigma$.}
\label{tab:batobservations}
\begin{center}
\begin{tabular}{l l l l l l}
Source  &$\sigma$& ct rate & flux & XRT-\\
XMMSL1\,J & & $\times$10$^{-5}$ & $\times$10$^{-5}$ & ~detected?\\
& & (ct cm$^{-2}$ s$^{-1}$) & (mCrab)& \\ \hline 
002202.9$+$254004 & 3.3 &7.8$\pm2.2$&  0.31& Y\\
044357.4$-$364413  & 3.1 & 4.6$\pm$1.9& 0.21 & N\\
093738.4$-$654445 & 3.0 & 7.0$\pm$2.1& 0.27 & N\\
125522.0$-$221035 & 3.2 &8.1$\pm$2.7&  0.36&Y \\
185608.5$-$430320                & 4.0   & 10.0$\pm$2.7      &      0.44&N \\
\end{tabular} 
\end{center}
\end{table}

\subsection{Optical and UV detections with \emph{Swift} UVOT}\label{sec:opt1}
A single UVOT source lies within the revised XRT error circle for the majority of the XRT-detected Slew Survey sources. We derived positions for and performed photometry on these sources using the {\it Swift} tool {\small uvotdetect}. The results are presented in Table \ref{tab:uvot}, where magnitudes are uncorrected for extinction and do not include any systematic uncertainties associated with the zeropoints \citep{Poole} but positional errors include a systematic uncertainty of 0.42$''$ \citep{Breeveld}. Twenty four of the XRT-detected sources are observed in the $b$ filter, and 16 sources have UV observations. There are a handful of very bright sources with $b<14$, while the mean $b$ magnitude is 16.4. Approximate limiting magnitudes for the UVOT for these sources are 20$<b<$11, consistent with the upper and lower bounds of the reported magnitudes. For completeness we include in Table \ref{tab:uvot} sources that we classified in Section \ref{rosatsec}. 
\setcounter{figure}{5}
\begin{figure*}
\begin{center}
\includegraphics[width=15cm, angle=0]{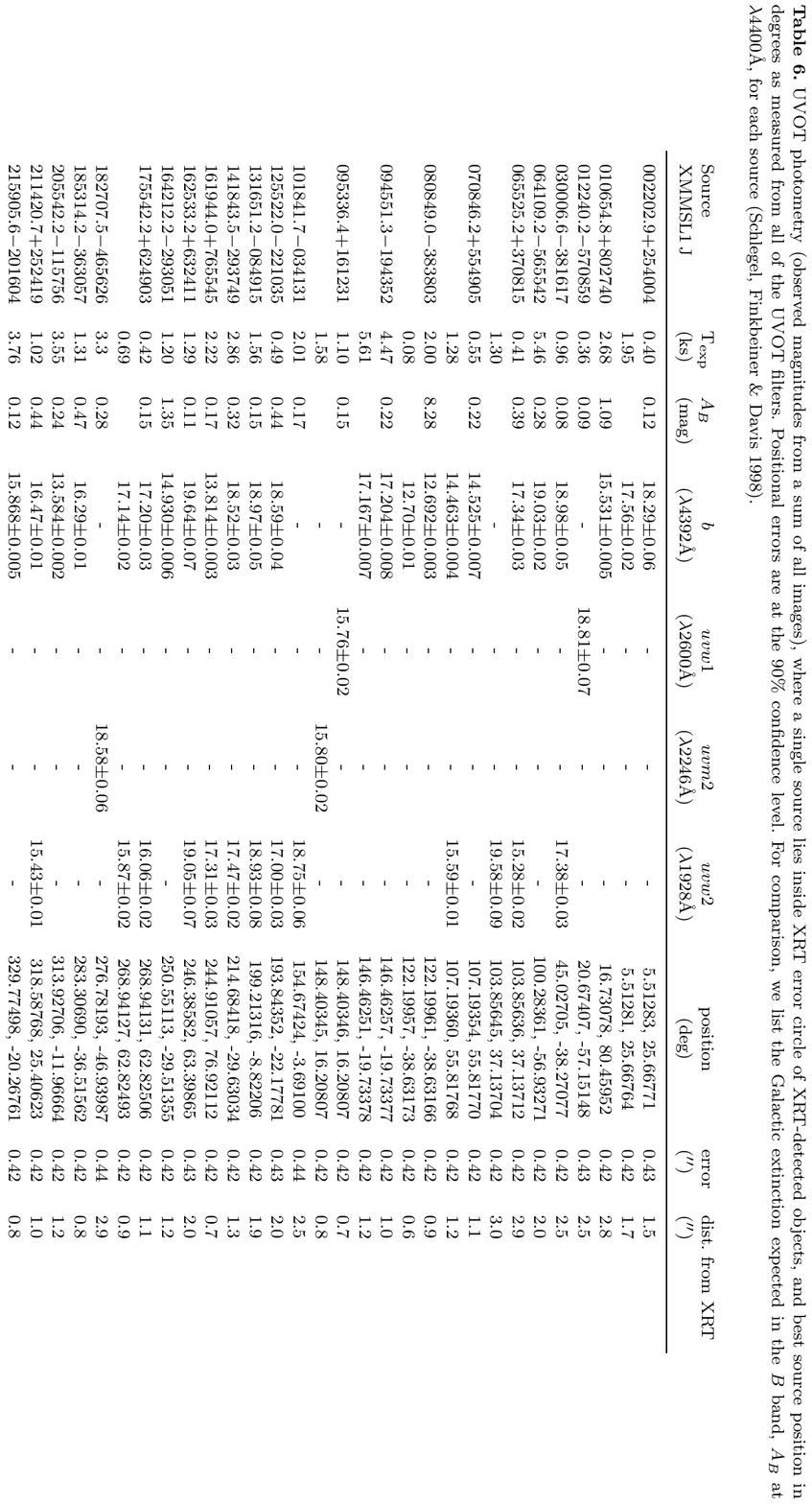}
\caption{}
\label{tab:uvot}
\end{center}
\end{figure*}

\setcounter{figure}{3}
\begin{figure}
\begin{center}
\includegraphics[width=6cm, angle=90]{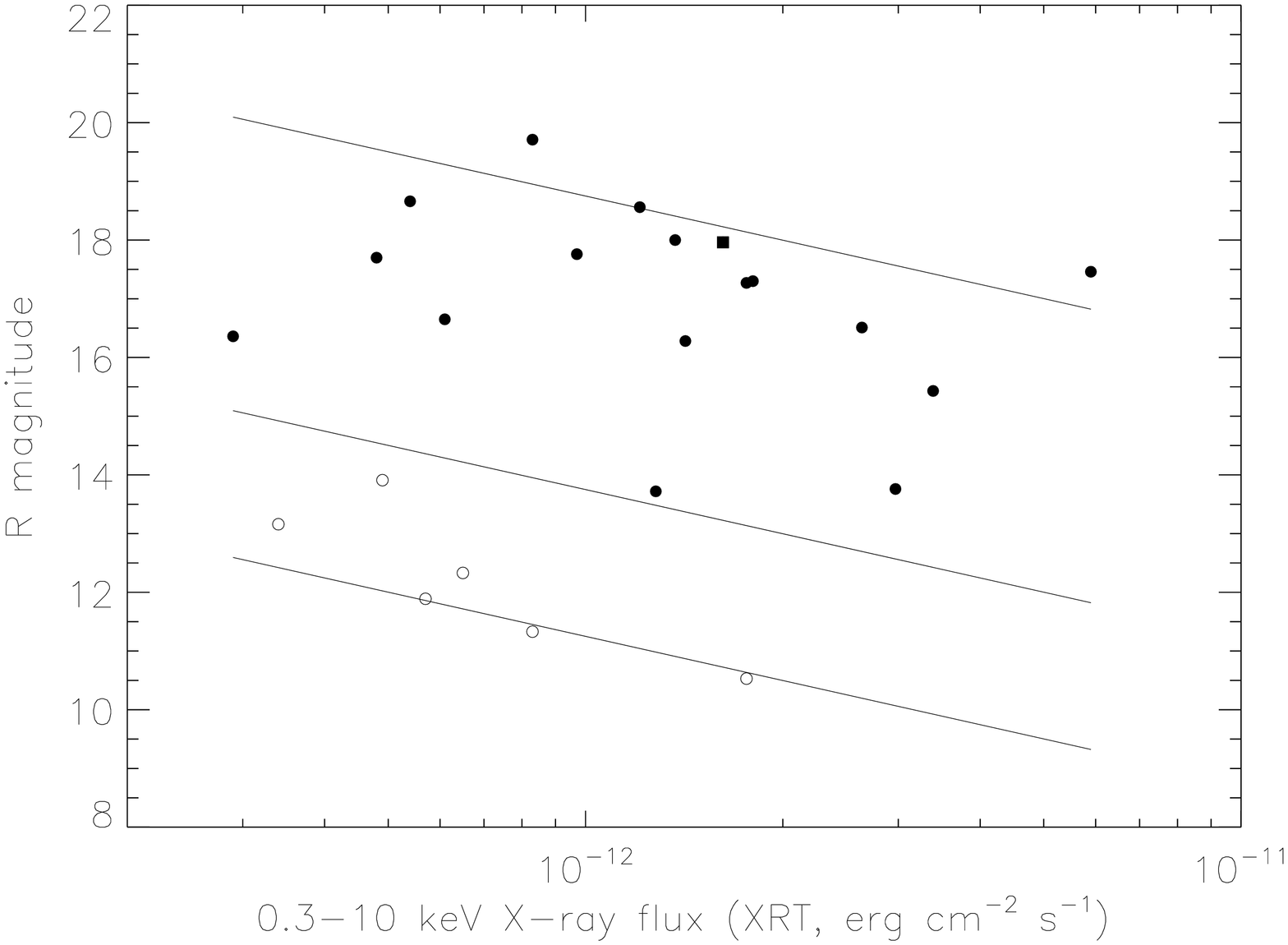}
\caption{X-ray to optical flux ratios for XRT-detected sources with potential catalogued $R$ band counterparts. The solid lines indicate $log \frac{F_X}{F_R}$ of 1 (upper line) and -1 (middle line), between which AGN are typically found (filled circles), and -2 (lower line) which extends this diagnostic to low luminosity AGN, starbursts and normal galaxies (open circles). The square symbol indicates a source with a $b \rightarrow B$ mismatch.}
\label{fxopt}
\end{center}
\end{figure}
\setcounter{figure}{4}
\begin{figure}
\begin{center}
\includegraphics[width=6cm, angle=90]{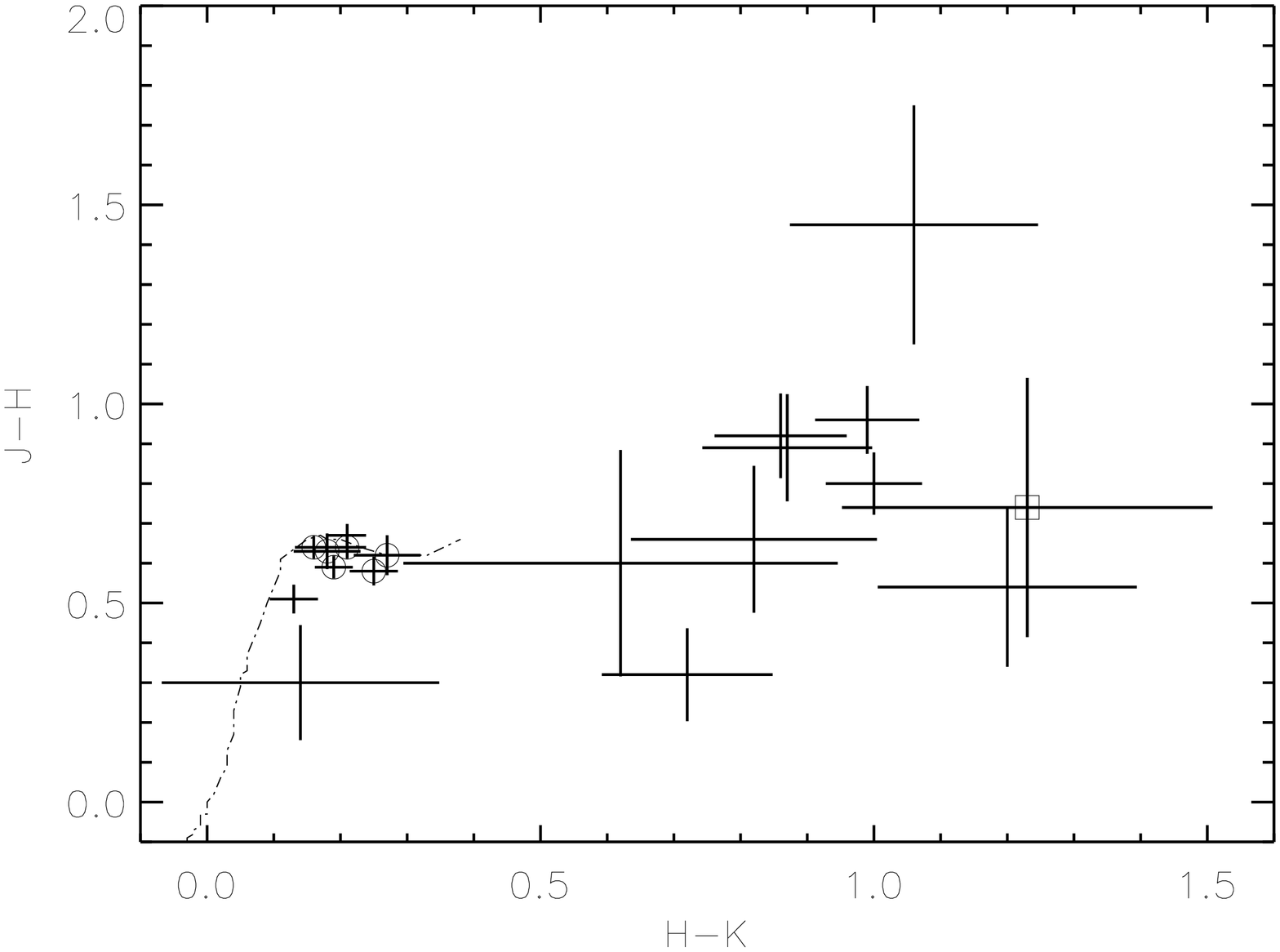}
\caption{2MASS colours for the objects which lie within XRT error circles, where only 1 unclassified source is listed. Magnitudes do not include any correction for extinction. The dot-dashed line indicates the expectation for A- through M-type main-sequence stars \citep{Bessell,Allen}. Open circles are overlaid on sources with $\log \frac{F_X}{F_R} < -1$ as depicted in Fig. \ref{fxopt}. The square symbol indicates a source with a $b \rightarrow B$ mismatch.} 
\label{colours}
\end{center}
\end{figure}

\subsection{Matches with catalogued optical and nIR sources}\label{sec:opt}
The UVOT spatial resolution is based on a point spread function full-width at half maximum (FWHM) of $\sim$2.2--3.0$''$, which might result in blends of two or more optical sources appearing as one object. To further pinpoint and characterise the optical sources found with UVOT we searched the Two Micron All Sky Survey \citep[2MASS][]{Cutri}, USNO-B1.0 \citep{Monet03}, USNO-A2.0 \citep{Monet98} and Naval Observatory Merged Astrometric Dataset \citep[NOMAD][including information from the unpublished USNO YB6 Catalog]{Zacharias} catalogues, via the {\small VizieR} search engine, to look for objects within the XRT error circles. This resulted in either single matches for which we report the $BVRIJHK$ magnitudes and any measured proper motions in Table \ref{tab:2mass}, or in three cases no single match was found. We compared the UVOT $b$ magnitude with the USNO.B1 $B$ magnitude for those sources for which both measurements are available. This comparison suggests that in the majority of cases the UVOT source and the catalogued source are likely to be the same object. The sources may be optically variable (quite likely given that many are X-ray variable) so this is only an indication of correspondence, but all sources with catalogued magnitudes of $B\sim 19$ or brighter are of similar magnitude to the UVOT detected source. At the faint end of the catalogued magnitude distribution we find two sources with UVOT counterparts two magnitudes or more brighter than the corresponding catalogued source. In these cases, either UVOT measures a different source, cannot resolve a blend of multiple sources, is detecting variability or one or both of the measured magnitudes are incorrect. 

Where there is a match with a single catalogued and identified source, we can further refine our sample. Four sources can be classified through this method, assuming that the catalogued match is the optical counterpart of the X-ray detected source. These are indicated in Table \ref{tab:2mass} and comprise one QSO at redshift $z = 0.87$, two high proper-motion stars one of which likely lies at d $<$ 33 pc and one variable star with a period of 0.7 d. We expect a number of flare stars to be included in this sample, given that they are both populous and highly variable.

The X-ray to optical flux ratio is often used as a method of classification of galaxies \citep[e.g.][and references therein]{Hornschemeier,Laird}. Here we calculated:
\begin{equation}
log(F_X/F_R) = log F_X + 5.5 + R/2.5,
\end{equation}
where $ F_X$ is the 0.3--10\,keV observed X-ray flux with XRT and $R$ is taken from the observed magnitudes in the USNO-B1.0 catalogue. The ratios we derive are listed in Table \ref{tab:2mass} and plotted in Fig. \ref{fxopt}: five sources can not be evaluated this way because 4 do not have $R$ band observations and 1 has no acceptable X-ray model fit. The log of the ratio of X-ray to optical fluxes results in values ranging from -2.05 to 1.30. AGN (both broad and narrow-line) typically show $-1~<~\log (F_X/F_R)~<~1$ \citep{Hornschemeier,Laird}, while lower values are obtained for low-luminosity AGN, starbursts and normal galaxies \citep[e.g.][]{Barger} and stars where an M star is expected to have $log (F_X/F_R) = -2$ \citep{Hornschemeier}. Of the 22 sources we tested in this sample, 15 have classic AGN-like ratios. Five sources have $log (F_X/F_R) < -1$, XMMSL1\,J080849.0$-$383803, XMMSL1\,J161944.0$+$765545, XMMSL1\,J101841.7$-$034131, XMMSL1\,J164212.2$-$293051 and XMMSL1\,J205542.2$-$115756. The first two of these we have classified as stars, while the remaining three have optical magnitudes among the brightest in the sample suggesting these may also be stellar in nature. Two sources have $log (F_X/F_R) > 1$, XMMSL1\,J002202.9$+$254004 and XMMSL1\,J030006.6$-$381617, and both of these have BAT detections possibly consistent with a contribution to the high energy emission from a jet (e.g. a blazar) though we note that magnetic CVs can also have large X-ray to optical flux ratios. 

We caution first that the X-ray fluxes we use are often based on low signal-to-noise spectra and a power law is not the only model that may provide a good fit to the data. Second, we have not corrected for the unknown amount of extinction and absorption along the line-of-sight to the sources so we are not calculating the intrinsic flux ratios. Third, we have to make the assumption that the catalogued $R$ band sources are indeed the optical counterparts to our X-ray detected sources and that no variability has occurred between the ground-based and {\it Swift} XRT observations.

Seven of the sources have substantial measured proper motions ($> 20$ or $<-20$ mas yr$^{-1}$), indicating that these must be relatively nearby stars.
In Fig. \ref{colours} we plot a colour-colour diagram using the 2MASS magnitudes and their errors for the catalogued sources, and compare these to the main sequence to see if any sources may be identified as stellar. We find that 9 sources lie along the main sequence for A- through M-type stars. None of the sources could be of earlier stellar type according to their nIR colours. Six of these 9 sources
have significant proper motions and 5 have X-ray to optical flux ratios lower than expected for AGN, supporting the classification of these as main-sequence stars. In fact, we find that two sources are associated with known stars from the coincidence of the UVOT and catalogued source positions (Table \ref{tab:2mass}). 

\setcounter{figure}{6}
\begin{figure*}
\begin{center}
\includegraphics[width=15cm, angle=0]{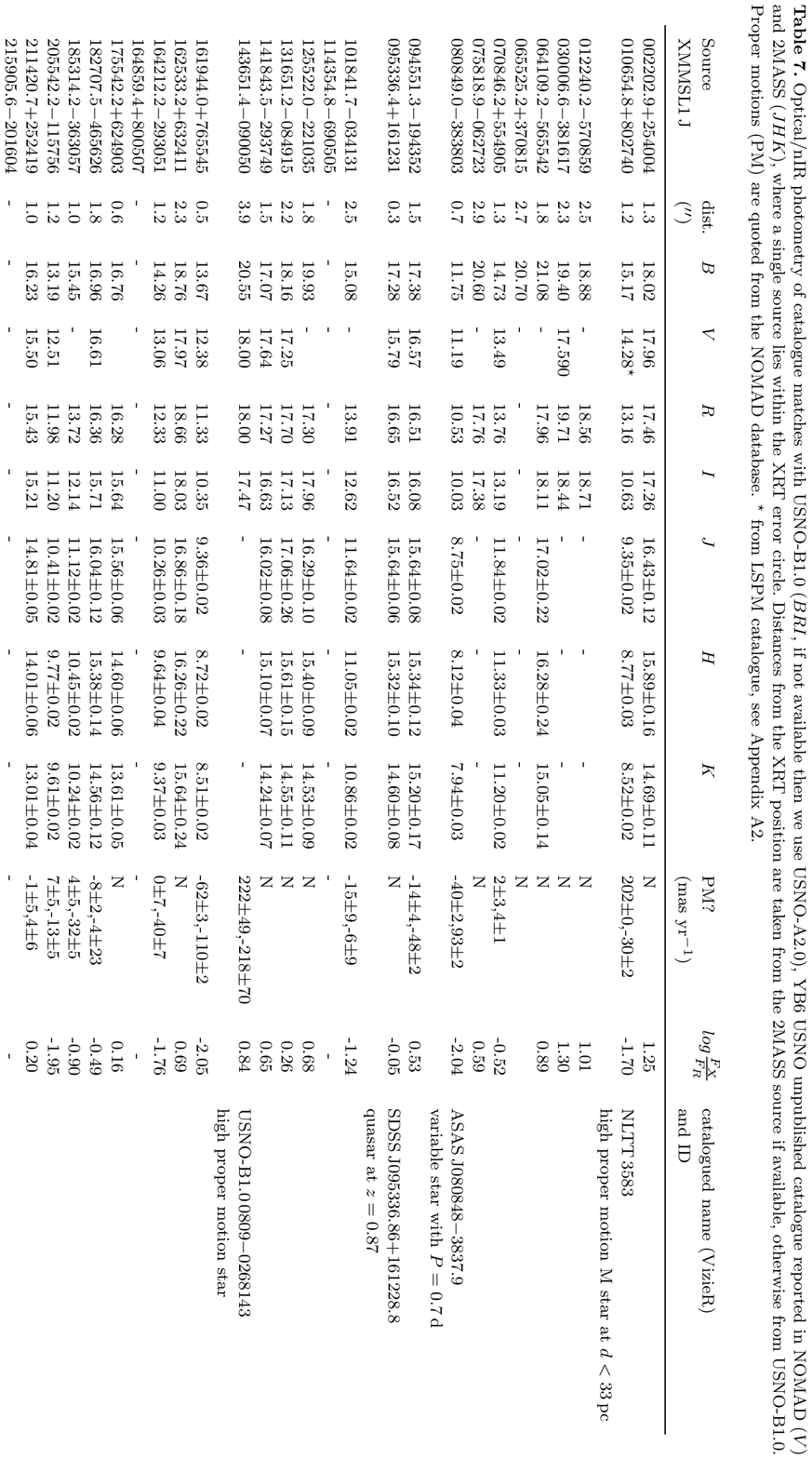}
\caption{}
\label{tab:2mass}
\end{center}
\end{figure*}

\setcounter{figure}{5}
\begin{figure}
\begin{center}
\includegraphics[width=7cm, angle=-90]{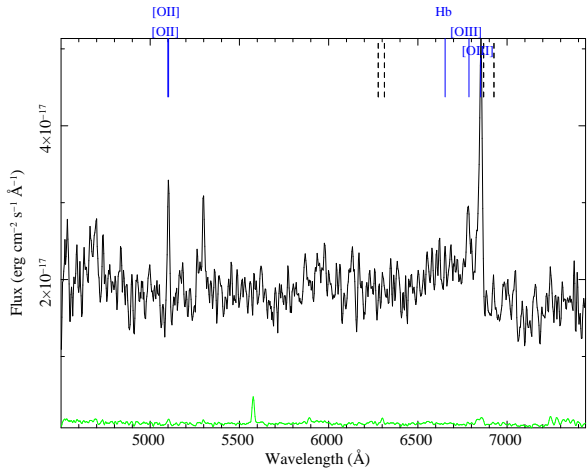}
\caption{NTT-EFOSC2 optical spectrum for XMMSL1\,J064109.2-56554 (black) and sky background spectrum underneath (grey), corrected for Galactic reddening, showing the optical emission lines used for the identification of the source. Atmospheric absorption lines are indicated with dashed lines.}
\label{spectrum}
\end{center}
\end{figure}
\subsubsection{Optical spectroscopy} \label{optspectroscopy}
An optical spectrum for XMMSL1\,J064109.2$-$565542 was obtained at the 3.6-m New Technology Telescope
(NTT) at La Silla, Chile on 2010 March 8. Two exposures of 600~s each were made with the ESO Faint Object
Spectrograph and Camera (EFOSC2) in good weather conditions, covering the wavelength range from $\sim$4500--7500\AA\ at 12\AA\ resolution. Details of the observational setup and data reduction are given in Appendix A5.
The object is classified as a broad-line AGN, based on the detection
of a broad (FWHM$>$1000 km s$^{-1}$) H$\beta$ 4861\AA\ line, at a redshift of $z=0.368\pm0.001$ (from the
detection of the [OII] $\lambda$3727, [OIII] $\lambda$4959
and [OIII] $\lambda$5007 emission lines) as shown in Figure \ref{spectrum}. We plan to establish or confirm the classifications presented in this work with optical spectroscopy of all previously unclassified optical counterparts to the XRT-detected sources.

\subsection{Matches with catalogued radio sources}\label{sec:radio}
We searched the radio catalogues available via VizieR on all the XRT- and BAT-detected sources to look for radio associations within 30$''$ of the XRT- and BAT-detected sources. Only one object, XMMSL1\,J164859.4$+$800507, has an associated radio source listed in the Atlas of Radio/X-ray associations \citep[ARXA,][]{Flesch}, the Westerbork Northern Sky Survey (WENSS, Leiden 1998) and the 1.4\,GHz NRAO VLA Sky Survey \citep[NVSS,][]{Condon}. The lack of an optical detection for this source with {\it Swift} UVOT is consistent with the approximate reported $B$ magnitude in the ARXA of 19.9. The ARXA also reports an $R$ magnitude of 16.2, implying that the source is extremely red perhaps due to dust extinction or high redshift, or it is highly variable since the $R$ and $B$ magnitudes were not necessarily obtained at the same epoch.

We then searched the VLA Faint Images of the Radio Sky at Twenty-cm catalogue \citep[FIRST,][]{White}, MIT-Green Bank 5-GHz Survey Catalog \citep[][and references therein]{Griffith}, Sydney University Molonglo Sky Survey \citep[SUMSS,][]{Mauch}, WENSS (Leiden 1998) and NVSS \citep{Condon} catalogues for radio emission within a few arcseconds of any of the {\it Swift}-detected sources. 
Again, only XMMSL1\,J164859.4$+$800507 has associated radio emission: NVSS\,164843+800516 lies 5.8$''$ away, within the XRT error circle, and has an integrated 20\,cm (1.4\,GHz) flux density of 3.8$\pm$0.5 mJy. Three further radio sources are located in this region, between 8$''$ and 1.5$'$ distant: WN\,1652.5+8009, WN\,1652.5$+$8009A and WN\,1652.5$+$8009B with 92\,cm (325\,MHz) flux densities spanning 30--80 ($\pm$2) mJy.

XMMSL1\,J164859.4$+$800507 has proven difficult to classify given its relatively poor positional uncertainty and lack of any clear optical counterpart, so the association with a radio source at 20\,cm will play a major role determining the nature of this object.
Interestingly, the candidate blazars XMMSL1\,J002202.9$+$254004 and XMMSL1\,J125522.0$-$221035, highly X-ray variable sources with large X-ray to optical flux ratios (the latter also detected with {\it Swift} BAT), are not associated with any known radio sources, within the scope of our search. 

\setcounter{table}{7}
\begin{table*}
\caption{Summary of proposed classifications for individual sources, and/or any associated catalogued sources.}
\label{tab:IDs}
\begin{center}
\begin{tabular}{l l l}
Source & proposed identification & associated catalogued sources\\\hline
XMMSL1\,J002202.9$+$254004 &AGN, possible blazar & \\
XMMSL1\,J010654.8$+$802740 &M-type star & NLTT\,3583 \\
XMMSL1\,J012240.2$-$570859 &AGN, possible NLS1 & 1RXS\,J012245.0$-$570901 \\
XMMSL1\,J030006.6$-$381617 &possible AGN & \\
XMMSL1\,J044357.4$-$364413 &possible AGN & \\
XMMSL1\,J063950.7$+$093634 & possible periodic M star, $P=1.36$ d & see Appendix A4\\  
XMMSL1\,J064109.2$-$565542 &Type I AGN, $z=0.368$ & 1RXS\,J064106.5$-$565610\\
XMMSL1\,J065525.2$+$370815 &possible QSO& \\
XMMSL1\,J080849.0$-$383803 &periodic variable star, $P=0.72$ d & ASAS\,J080848$-$3837.9\\
XMMSL1\,J093738.4$-$654445& candidate Galactic hard X-ray flash& \\
XMMSL1\,J094551.3$-$194352 & flare star& \\
XMMSL1\,J095336.4$+$161231& QSO, $z=0.87$ &SDSS\,J095336.86$+$161228.8 \\
XMMSL1\,J101841.7$-$034131& M star& \\ 
XMMSL1\,J114354.8$-$690505& -&multiple {\it ROSAT} matches \\ 
XMMSL1\,J125522.0$-$221035&AGN, possible blazar & \\ 
XMMSL1\,J131651.2$-$084915& AGN& \\ 
XMMSL1\,J141843.5$-$293749&AGN& 1RXS J141846.1$-$293748\\  
XMMSL1\,J143651.4$-$090050&high proper motion star& USNO-B1.0\,0809$-$0268143 / 1RXS\,J143653.7$-$090004\\  
XMMSL1\,J161944.0$+$765545& late-type m-s star&XMMSL1\,J161935.7$+$765508 / 1RXS\,J161939.9$+$765515\\ 
XMMSL1\,J162533.2$+$632411 &AGN, possible Type II & \\
XMMSL1\,J164212.2$-$293051 &M star&1RXS\,J164216.5$-$293035\\ 
XMMSL1\,J164859.4$+$800507&possible AGN & 1RXS\,J164843.5$+$800506 / NVSS\,164843$+$800516\\ 
XMMSL1\,J175542.2$+$624903&Type I AGN, $z=0.236$ &1RXS\,J175546.2$+$624927 \\ 
XMMSL1\,J182707.5$-$465626 & possible AGN& \\
XMMSL1\,J185314.2$-$363057 &possible M star& \\
XMMSL1\,J185608.5$-$430320& possible AGN& \\
XMMSL1\,J205542.2$-$115756 &K-M type star& \\ 
XMMSL1\,J211420.7$+$252419 &possible AGN& \\ 
\end{tabular} 
\end{center}
\end{table*}

\setcounter{figure}{6}
\begin{figure}
\begin{center}
\includegraphics[width=8cm, angle=0]{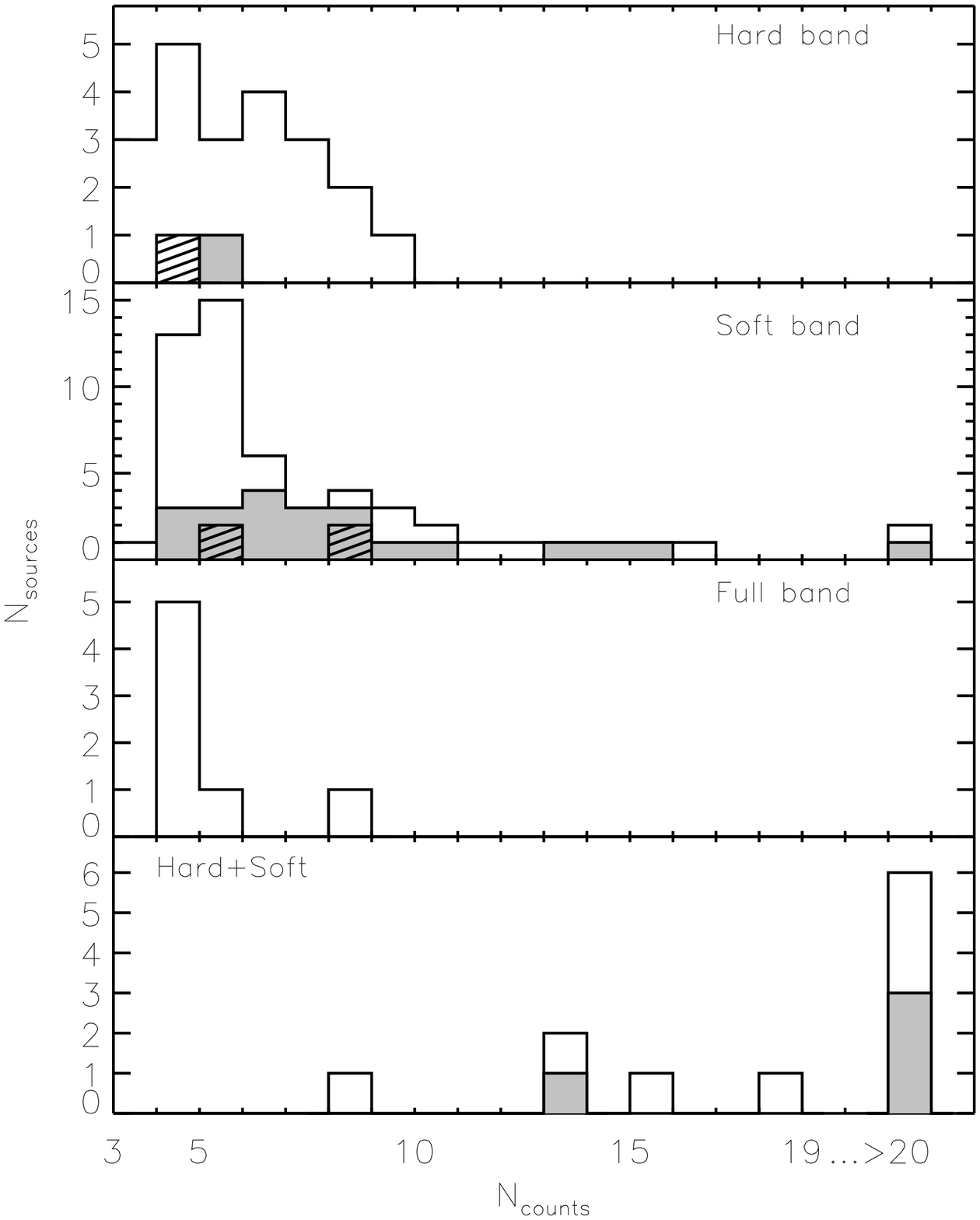}
\caption{The numbers of BAT/XRT-detected (grey shading for XRT detections, hatching for BAT detections) and non-detected sample sources against the number of measured counts reported in the {\it XMM-Newton} Slew Survey observation. The four panels show the distribution over the Slew Survey energy bands: hard or hard+full band (2--12 keV, top), soft or soft+full band (0.2--2 keV, upper middle), full band only (0.2--12 keV, lower middle) and hard+soft+full bands (lower panel).}
\label{slewhistograms}
\end{center}
\end{figure}

\section{XRT-undetected sources} \label{sec:undetected}
The majority of our sample of unidentified X-ray sources from the {\it XMM-Newton} Slew Survey are not detected with {\it Swift}. While the nature of these sources is difficult to determine, it is very important to attempt to do so in order to understand the complete population identified by the Slew Survey. We expect to pick up large numbers of transients due to the requirement that sample sources were not already identified from a variety of multiwavelength catalogue searches (Section \ref{sample}). We only required a detection in one of the three Slew Survey energy bands, with a minimum of 4 source counts, so we also expect that some fraction of these identifications will be spurious.

Firstly, we address the issue of spurious detections. Statistical considerations leads to an estimate of 4\% that are expected to be spurious sources in the {\it XMM-Newton} Slew Survey clean catalogue \citep{Saxton}. This figure does not apply directly to our sample, as it was calculated on full-band detections and for the entire catalogue (i.e. without applying the source selection criteria used here), but can be used as a guide. The {\it XMM-Newton} full-band detection likelihoods of our sample range from 10 to 556 \citep[for details see][]{Saxton}. The mean detection likelihoods for the XRT-detected and XRT non-detected sources are 37 and 32 respectively, i.e. the confidence in these detections is approximately equal for an average source from each population, using this measure alone.

We searched the 2MASS catalogue for bright nIR sources coincident with the {\it XMM-Newton} Slew Survey positions (the nIR being less susceptible to extinction than optical bands). We found that 27\% of the XRT non-detected Slew Survey sources have one or more catalogued nIR sources brighter than $J = 14$ mag within 15$''$. To estimate the fraction for which this will be true simply by chance, we generated 10000 random sky positions and cross-correlated these with the 2MASS catalogue. This leads to the expectation that a 2MASS source brighter than $J=14$ mag will appear within 15$''$ of a random sky position around 7.9\% of the time. The fraction expected by chance is therefore much lower than obtained for our sample of XRT non-detected sources. We can conclude from this that some fraction of our non-detected subset must be real. For our XRT-detected subset, only 30\% of the sources would have been recovered when performing the same $J<14$ mag cut; for example in the case of flare stars we expect only the closest sources to be found this way. So finding a bright nIR counterpart strongly suggests the X-ray source is real, while the lack of a counterpart is not constraining.

We then investigated the distribution of our detected and non-detected subsets over the Slew Survey energy bands. Sources can appear in any one of the hard (2--12 keV), soft (0.2--2 keV) and full (0.2--12 keV) bands, or could appear in multiple bands. Sources which appear in all three bands we term hard+soft. Figure \ref{slewhistograms} shows the numbers of detected (shaded areas) and non-detected sample sources that are hard (only), soft (only), full (only) or hard+soft band detections in the Slew Survey against the number of measured counts per source. We see that most of the sample were soft only (soft or soft+full band) Slew Survey detections, reflecting the observed X-ray source population in general. We also expect a greater number of objects with low numbers of counts in the Slew Survey soft band than in the hard band because, a) many of these sources are expected to be unabsorbed AGN: their emission is greater in the soft band and follows a power law such that sources close to the detection limit may appear soft-only, and, b) the background is higher in the hard band making source detection more difficult. The bright nIR counterpart search described above returned a match for XRT non-detected sources from the hard, soft and hard+soft categories, in the ratios 1.0~:~2.5~:~0.2 (for comparison, all XRT non-detected sources follow the ratios 1.0~:~1.65~:~0.2 for the same bands). 

There are 11 sources detected in all 3 Slew Survey bands, and we can be confident that these sources are all astrophysical yet only 36\% are Swift-detected. This suggests that a significant fraction of the XRT non-detected sample are highly variable sources.
Of the 55 soft-only sources, 22 are XRT detected, a further 2 are BAT detected and a further 12 have
$J<14$ mag nIR sources within 15$''$. From our estimates, described earlier, $\sim$2 of the soft-only sources may be coincident with a $J<14$ mag nIR source by chance. We can therefore say that at least 34 out of 55 (62\%) are likely to be real astrophysical sources. The remaining 38\% may comprise 30\% with $J$ band counterparts below the brightness cut we imposed, following the proportion seen in the XRT-detected source population; just 8\% or approximately 4 sources are then left unaccounted for and could potentially be spurious.
Of the 21 hard-only sources, just one is detected with XRT and one is detected with BAT. A further 5 have
bright nIR sources in the vicinity of which 1--2 may be chance coincidences. If we again adopt the fraction 30\% for real sources likely to have nIR counterparts fainter than $J=14$, we find 38\% of all the XRT/BAT non-detected hard-only sources would be considered real. This provides an upper bound on the fraction of hard-only sample sources that are potentially spurious. 
Full band only Slew Survey sources comprise 10\% of the XRT non-detected sources. We have found no catalogued bright, nearby $J$ band sources for this subsample. 
In summary, there are likely to be some spurious sources included in the full-only and hard-only subsamples, while the vast majority if not all the soft-only sources as well as all the hard+soft sources are probably astrophysical.

We now focus on the nature of the XRT non-detected sources.
The distribution on the sky of XRT non-detected sources appears very similar to that of the detected sources (Fig. \ref{aitoff}). The median of the Galactic column densities, $N_{\rm H,Gal}$, in those directions is not significantly different for the detected and non-detected samples, but we note that this could have hampered the detection of XMMSL1\,J164456.7$-$450015 and XMMSL1\,J183233.0$-$112539 which may lie behind particularly large Galactic columns of 1.65$\times$10$^{22}$ cm$^{-2}$ and 1.13$\times$10$^{22}$ cm$^{-2}$ respectively. The lack of detections with {\it Swift} BAT argues against a population of heavily obscured AGN (Section \ref{bat}). To support this we compared expected XRT count rates for typical AGN at $z=0.1$ with observed 0.3--10\,keV flux 10$^{-12}$ erg cm$^{-2}$ s$^{-1}$, power law photon index $\Gamma=1.7$, Galactic column density $N_{\rm H} = 10^{20}$ cm$^{-2}$ and intrinsic X-ray absorbing columns of $N_{\rm H} = 10^{21}$, 10$^{22}$ and 10$^{23}$ cm$^{-2}$, and find the count rate is reduced by 64\% at most, which is not enough to push these sources into our XRT non-detected category. An intrinsic column density of $N_{\rm H} = 10^{24}$ cm$^{-2}$ is required to bring the 0.3--10\,keV XRT count rate for this spectral shape down to the mean detection limit for this sample.

We require greater variability among the non-detected sources, shown in Fig. \ref{countrates}a; with variability corresponding to flux changes of up to a factor $\sim$300. These highly variable sources remain an enigma and are likely to do so until they are seen again, perhaps as they undergo an X-ray outburst or otherwise enter a high flux state.

\section{Discussion and conclusions}
{\it Swift} observations of a sample of 94 unidentified X-ray sources from the {\it XMM-Newton} Slew Survey have been carried out, with 29\% of the sample sources detected with XRT. This low detection rate supports the hypothesis that many of these sources are highly variable or X-ray transient objects. The X-ray emission or upper limit to the emission for all the sources, taking into account count rate conversion between instruments and Eddington Bias, lies at or below that seen in the Slew Survey. Up to two thirds of the XRT detected sources could have remained constant in flux between the Slew Survey and {\it Swift} observations. Approximately one third of the XRT detected sources and also the majority of the XRT non-detected sources are likely to be variable. 

The X-ray positions we derived from the {\it Swift} data for the XRT-detected sources improved the mean 90\% confidence error radius from 18.9$''$ to 2.9$''$. This reduced the number of UVOT and catalogued optical matches to just a single source in most cases. Performing a new cross-correlation of the 3$\sigma$ error radii with multiwavelength catalogues revealed that six sources can be associated with known objects and 8 sources may be associated with unidentified {\it ROSAT} sources. The X-ray spectrum of most of the sources can be fit with an absorbed power law with photon indices clustering around $\Gamma = 1.5 - 2.0$, typical of AGN. The random distribution across the sky of this and the non-detected population is also consistent with an AGN classification. To identify the types of objects included in the XRT detected sample we used a number of further indicators: the X-ray to optical flux ratio, proper motion, nIR colours, radio associations and detection in $\gamma$-rays with BAT. We summarise the proposed source classifications in Table \ref{tab:IDs}.
We find 10 of the 30 XRT- and/or BAT-detected sources are clearly stellar in nature, including one periodic variable star and 2 high proper motion stars. Eleven sources are classified as AGN, 4 of which are detected in hard X-rays with BAT and 3 of which have redshifts spanning $z = 0.2 - 0.9$ obtained from the literature or from optical spectroscopy. A further 3 sources are suspected AGN and 1 is a candidate Galactic hard X-ray flash, while 5 sources remain unclassified. Interestingly, the 2 most variable sources on time scales of a few years (between the {\it XMM-Newton} and {\it Swift} observations) are among those which we cannot classify here. We plan to obtain optical spectroscopy where possible for all these sources, which will confirm or determine their identifications, as demonstrated in Section \ref{optspectroscopy} for XMMSL1\,J064109.2$-$565542.

The XRT/BAT non-detected population are equally important to classify, but the lack of information makes this task far more difficult. The majority of these are likely highly variable sources, and from the lack of BAT detections we can to a large extent rule out a population of heavily obscured AGN. We also expect some fraction of these sources may be spurious detections. 
We compared the non-detected population with the detected population in terms of the distribution of {\it XMM-Newton} Slew Survey counts in each of the hard, soft and full bands. The X-ray error circles from the Slew Survey are somewhat large for a full optical/nIR counterpart search, so instead we looked for bright nIR sources within 15$''$ of the Slew Survey position, and compared this to the number expected in a chance coincidence. Combining these two sets of information, we estimate the fraction of astrophysical sources as opposed to spurious detections among the XRT non-detected population for each Slew Survey band. All the hard+soft band detections are extremely likely to be real, and we find that most if not all the soft sources are also likely to be real; 73\% of all sample sources fall into these two categories. Perhaps 60\% of the 21 hard-only sources could be spurious. We stress, however, that these figures are only estimates and there is potential for as yet unknown source types within these populations. It is likely that the nature of each {\it Swift} non-detected source will remain elusive until they are once again detected, permitting further study. 

In summary, the XRT-detected population seems to consist of approximately equal numbers of X-ray active stars and background AGN, while the undetected population may contain more extragalactic objects such as AGN. Type II AGN were perhaps expected to be the dominant population due to their lack of soft X-ray emission, given that this sample was selected based on {\it ROSAT} soft X-ray non-detections, but we identify only one possible Type II candidate among the XRT-detected population. Neither are they numerous among the XRT non-detected sources as implied by the lack of BAT detections. A knowledge of the source types detected in surveys such as the {\it XMM-Newton} Slew Survey is important for investigation of the $\log N - \log S$ and completing studies of the X-ray background that cannot be done with pointed observations alone. Follow-up of the Slew Survey sources with {\it Swift} has also enabled the identification of a highly variable population, largely of unknown nature.

\section{Acknowledgments}
RLCS, PAE, AMR, PE and JPO acknowledge financial support from STFC. We thank the {\it Swift} team for performing these observations. We acknowledge useful discussions with S. Farrell and M. Goad and the anonymous referee for a constructive critique. This work made use of data
supplied by the UK {\it Swift} Science Data Centre at the University of Leicester. This publication makes use of data products from the Two Micron All Sky Survey, which is a joint project of the University of Massachusetts and the Infrared Processing and Analysis Center/California Institute of Technology, funded by the National Aeronautics and Space Administration and the National Science Foundation. Based on observations obtained at the ESO NTT telescope (programme 084.A-0828).

\appendix
\section{Notes on individual sources}
\subsection{XMMSL1\,J002202.9+254004}
Two {\it Swift} pointed observations were taken 10 months apart, during which we see significant variability in both the X-ray (count rate increasing by a factor of $\sim$5) and optical (an increase of 1 magnitude) bands. This, together with a high X-ray to optical flux ratio, suggest this is a blazar whose X-ray emission is boosted by contributions from a jet, while the lack of a radio association casts doubt on this classification.

\subsection{ XMMSL1\,J010654.8+802740}
This source is detected with the XRT with one of the lowest count rates of the sample though Galactic absorption in this direction is very high. The signal to noise in the X-ray spectrum is very low and we cannot model the spectral shape but we find that this source is X-ray variable on time scales of a few years. A bright UVOT counterpart is present despite having the second highest reddening among the detected sources equating to 1.09 magnitudes in $B$ band. 
We note that the $V$ magnitude listed in NOMAD for this source, $V = 17.8$, is inconsistent with the LSPM-North proper-motion catalog of nearby stars, \citep{Lepine} listing of $V = 14.28$ and the latter is a more reasonable extrapolation of catalogued magnitudes in other wavebands. 

\subsection{ XMMSL1\,J012240.2-570859}
If this source is an AGN, the blue colour and soft X-ray spectrum may indicate a large soft X-ray excess and a strong Big Blue Bump suggestive of a narrow-line Seyfert I galaxy.

\subsection{ XMMSL1\,J063950.7+093634}
This source is not detected with {\it Swift}. We note the presence of a $J\sim14$ M-type variable star in the galaxy NGC\,2264 with period 1.36 d at 5$''$ distance from the {\it XMM-Newton} Slew Survey position, within the error circle \citep{Rebull,Lamm}.

\subsection{ XMMSL1\,J064109.2-565542}
An optical spectrum for this source was obtained at the NTT, La Silla (Chile) on 2010 March 8.
Observing conditions were good with sky transparency clear to photometric and a
seeing of $\sim$1$''$.
Two exposures of 600~s each at an airmass of 1.1 were made with the ESO Faint Object
Spectrograph and Camera (EFOSC2) with a 1$''$ slit width
oriented at the parallactic angle and grating 4, covering the wavelength range from $\sim$4500--7500\AA\ at 12\AA\ resolution (from unblended arc
lines taken through the slit at $\sim 6000$\AA). A standard reduction process was applied using {\sc IRAF}
routines. Wavelength calibration was carried out
by comparison with exposures of an Helium-Argon arc lamp, with an accuracy $<1$\AA. Relative flux
calibration was carried out by observations
of the spectrophotometric standard star LTT 3218 \citep{Hamuy}. We estimate an error on the flux calibration $<$10\% from the standard adjustment
during the calibration procedure. The spectrum has been corrected for Galactic reddening. The object is classified as a broad-line AGN, based on the detection
of a broad (FWHM$>$1000 km s$^-1$) H$\beta$ 4861\AA\ line, at a redshift of 0.368$\pm$0.001 (from the detection of the [OII] $\lambda$3727, [OIII] $\lambda$4959
and [OIII] $\lambda$5007 emission lines) as shown in Figure \ref{spectrum}.

\subsection{ XMMSL1\,J065525.2+370815} 
Using the approximate X-ray flux from the power law spectral fit, we compared this to the flux during the {\it XMM-Newton} Slew Survey observation and find significant variability of a factor of at least 20. Consistent with the position of the UVOT counterpart we find an optical counterpart listed in USNO-B1.0 of order 3 magnitudes fainter in $B$ than the UVOT source. This source is also listed in SDSS DR7, where its magnitude is approximately the same as derived from the UVOT observations. This source is therefore variable in both X-ray and optical wavebands.

\subsection{ XMMSL1\,J070846.2+554905}
The XRT flux is among the highest in the sample, and the spectrum is well fitted with a power law of photon index 1.8, typical of AGN, while the absorption is non-zero and consistent with the Galactic absorption in that direction. A very small proper motion is reported in the catalogues searched, and the nIR colours are consistent with a K-type main sequence star. However, the X-ray to optical flux ratio is consistent with an AGN. This source will remain unclassified until optical spectroscopy can be obtained.

\subsection{ XMMSL1\,J075818.9-062723}
The UVOT $b$ and USNO $B$ magnitudes for this source differ by 2.5 magnitudes, indicating either large variability or inaccurate photometry. The X-ray spectrum is seen to harden between the {\it XMM-Newton} and {\it Swift} observations, being formally detected only in the soft band with {\it XMM-Newton} but seen in the full {\it Swift} energy band. The X-ray to optical flux ratio lies right in the middle of the expected range for AGN while the absence of any nIR emission which is uncommon for an AGN. 
This source requires further observations in order to determine its nature.

\subsection{ XMMSL1\,J080849.0-383803}
This source is very likely Galactic because the X-ray column density measured in a spectral fit (0.15$\pm$0.10~$\times$~10$^{22}$ cm$^{-2}$) is lower than the Galactic column in that direction of 0.78~$\times$~10$^{22}$ cm$^{-2}$ \citep{Kalberla}. The LAB HI maps show that this high Galactic column exists both at the nearest measured position to the XRT position, 0.06 degrees away and is the result with weighted interpolation over all the nearest measured values within a 1 degree radius. The LAB Survey is the most sensitive Milky Way HI survey to date, with the most extensive coverage both spatially and kinematically. 
We extracted the X-ray flux seen by RASS at the new XRT enhanced X-ray position (Table \ref{tab:positions}) and using the spectral shape measured by XRT (Table \ref{tab:spec}) and recover a detection. The X-ray flux appears to have decreased by 60\% over 17.7 years.
The UVOT-enhanced X-ray position we derive, with error radius 1.7$''$ (90\% containment) corresponds to a bright UVOT source of $b$ magnitude 12.69.
Its catalogued optical and nIR colours match those of a late K star, and this source has a proper motion typical of a thin disk star. The X-ray to optical flux ratio also suggests a stellar nature for this source. The position coincides with that of ASAS J080848-3837.9, a variable star listed in the AAVSO International Variable Star Index VSX \citep{Watson} of unknown type but with a period of 0.72 days.

\subsection{ XMMSL1\,J090822.3-643749}
This source is not detected with {\it Swift} despite showing the highest full-band {\it XMM-Newton} pn count rate of the entire sample of 9.9$\pm$0.9 count s$^{-1}$ (most of the counts fell in the soft band). The XRT detection limit was 0.002 count s$^{-1}$, indicating a factor of at least 100 decrease in X-ray flux between the {\it XMM-Newton} observations in 2004 and the {\it Swift} observations in 2006 at flux levels of $\sim$3$\times$10$^{-11}$ to $<1\times$10$^{-13}$ erg cm$^{-2}$ s$^{-1}$ respectively (assuming a typical AGN spectrum). {\it XMM-Newton} slewed over this position on two further occasions, in 2002 and 2008, during which the source was not detected to full-band limits of $\le$0.4 and $\le$0.3 count s$^{-1}$ respectively.

\subsection{ XMMSL1\,J094156.1+163246}
This source is not detected with {\it Swift}. We note, however, that a Sloan Digital Sky Survey (SDSS) galaxy with measured redshift of $z=0.17$ lies within the 3$\sigma$ error circle at 40$''$ radius. 

 \subsection{XMMSL1\,J094551.3-194352}
This source is detected with XRT and UVOT. The X-ray position has been determined at the most accurate level achieved for this sample, with a 90\% error radius of 1.5$''$. Multiple {\it Swift} observations show variability, decreasing by a factor of 1.4 in soft X-ray flux on time scales of days, and the spectrum has hardened in the 2.2 years between the {\it XMM-Newton} and XRT observations. We extracted the X-ray flux seen by RASS at the new XRT enhanced X-ray position (Table \ref{tab:positions}) and using the spectral shape measured by XRT (Table \ref{tab:spec}) and recover a detection. A comparison of the 0.3--2\,keV XRT flux with the RASS flux at the XRT position shows significant variability: a factor of 4.2$^{+0.6}_{-0.8}$ (1$\sigma$ error) increase over 18 years. 

\subsection{ XMMSL1\,J095336.4+161231}
This source is detected with XRT and UVOT. The X-ray spectral shape is poorly constrained, but the X-ray and optical fluxes are of the same order suggesting this is an AGN. At the XRT position there is a known quasar: SDSS\,J095336.86$+$161228.8 which lies at a redshift of $z = 0.87$ \citep{Adelman}. The XRT observations span 1 year, during which time the source halved in X-ray flux.

\subsection{ XMMSL1\,J114354.8-690505}
The X-ray flux at the time of the XRT observation of this source is approximately the same as measured with {\it XMM-Newton} and twice that measured at the XRT position with {\it ROSAT}, 5 and 18.6 years previously respectively.
Three {\it ROSAT} HRI sources lie within the 3$\sigma$ XRT error circle, listed in Table \ref{tab:rosat}. The {\it ROSAT} reported count rates for all of these sources are almost identical: 0.029$\pm$0.005, 0.028$\pm$0.005 and 0.027$\pm$0.005 count s$^{-1}$ respectively, and it is not clear which object, if indeed these are distinct objects, corresponds to the XRT- and {\it XMM-Newton}-detected X-ray source. Given the lack of optical and radio information it is difficult to classify this source.

\subsection{ XMMSL1\,J125522.0-221035}
The X-ray source was observed twice with {\it XMM-Newton}, both times only formally detected in the soft band. Comparison of the X-ray flux as observed with XRT, {\it XMM-Newton} and {\it ROSAT} reveals variability: the source is at least three times brighter in XRT than in {\it ROSAT} observations 17 years earlier (in the {\it ROSAT} pass band) and doubled in flux in the 2.4 years between the two {\it XMM-Newton} observations. The $b$ band magnitude measured with UVOT is 1.5 magnitudes brighter than the catalogued $B$ value, again indicating variability. All indicators used in this study point to an extragalactic source.

 \subsection{XMMSL1\,J131651.2-084915}
The UVOT magnitudes lie at the faint end of the UVOT detectability range, while the longer wavelength catalogued magnitudes at the same position are somewhat higher and show that this is a red object. The XRT X-ray flux has decreased to a tenth of the {\it XMM-Newton}-observed value. We extracted the X-ray flux seen by RASS at the new XRT enhanced X-ray position (Table \ref{tab:positions}) using the spectral shape measured by XRT (Table \ref{tab:spec}) and recover a detection. No variability is detected between the XRT and RASS observations 16.4 years apart. We note the optical flux and spectral shape are very similar to that of XMMSL1\,J141843.5-293749, which we suggest is an AGN, while the X-ray flux is 3 times lower.

 \subsection{XMMSL1\,J141843.5-293749}
The XRT position for this source has the best accuracy achieved in this sample, with an error radius of 1.5$''$ (90\% containment). The X-ray flux during the XRT observation was at least two times lower than observed with {\it XMM-Newton} but approximately the same as that of the associated {\it ROSAT} source.

\subsection{XMMSL1\,J143651.4-090050}
This source, classified here as a high proper motion star, is $\sim$12$''$ from a galaxy cluster at $z = 0.08$ \citep{Boehringer,Jones}.

\subsection{XMMSL1\,J161944.0+765545}
This source is detected with XRT and UVOT. The XRT refined position is coincident both with the {\it ROSAT} source 1RXS\,J161939.9$+$765515 and with the {\it XMM-Newton} Slew Survey source XMMSL1\,J161935.7$+$765508. The Slew Survey source 161935.7$+$765508 is not part of our sample due to its association with 1RXS\,J161939.9$+$765515. In the $\sim$17$\times$17 arcmin field of view of XRT only one X-ray source is detected, and its UVOT-enhanced X-ray position (with a 90\%  error radius of 1.8$''$), lies 13$''$ from the {\it XMM-Newton} position of 161939.9$+$765515 (90\%  error radius of 18$''$) and 34$''$ from that of 161944.0$+$765545 (90\%  error radius of 22$''$). Within their 3$\sigma$ error circles these 3 positions are all consistent, and are most probably one and the same source. The Hamburg/RASS Catalog of optical identifications V3.0 \citep{Zickgraf} provides an F-G star classification for the bright optical counterpart to the RASS source, 2MASS\,J16193872$+$7655165, which lies within the XRT error circle. The {\it Swift} XRT X-ray spectrum is not well fitted with an absorbed power law, requiring a very soft photon index of $\Gamma\sim4.6$ and a high absorbing column of order 4$\times$10$^{22}$ cm$^{-2}$ (10 times higher than the Galactic column). We performed, instead, an absorbed mekal fit to these data, giving a plasma temperature of $kT = 1.2^{+0.1}_{-0.2}$\,keV and an upper limit on the total column density lower than the mean Galactic column, suggesting this source is located between us and the far side of our Galaxy. The X-ray to optical flux ratio we measure places this source outside the region of typical AGN, and the nIR colours show that it is consistent with being a late-type main sequence star as reported by \cite{Zickgraf}.

 \subsection{XMMSL1\,J162136.0+093304}
This source is not detected with {\it Swift}. We note that this field has been observed with {\it Swift} for 7.09\,ks, with no detection to a deep 3$\sigma$ upper limit of 8.2$\times$10$^{-4}$ count s$^{-1}$. 

 \subsection{XMMSL1\,J162533.2+632411}
This source is detected with XRT and has a faint UVOT counterpart. The X-ray spectral parameters are difficult to constrain, however this is spectrally the hardest detected XRT and was detected only in the full band and hard bands in the {\it XMM-Newton} Slew Survey. 
We see variability in the X-ray flux of at least a factor of three on a time scale of five years. The X-ray position is consistent with {\it ROSAT} source 1RXS\,J162535.1$+$632333: given the hard spectrum of this source in {\it XMM-Newton} and in {\it Swift} observations this identification is uncertain. We therefore searched for a source in RASS at the new XRT enhanced X-ray position (Table \ref{tab:positions}) which resulted in a non-detection. This source may be a variable, absorbed AGN.

\subsection{XMMSL1\,J164859.4+800507}
The XRT position, the most uncertain of this sample, coincides with the {\it ROSAT} source 1RXS\,J164843.5$+$800506, which could not be classified as it is reportedly blended with another source in the Hamburg/RASS data. No soft X-ray variability is detected between the XRT and RASS observations (using now the XRT position to extract a RASS flux) 16.5 years apart. Our catalogue searches resulted in two optical sources within the error circle: a $\sim$14 mag source 1$''$ distant and a $\sim$19 mag source 3.4$''$ distant. This source is clearly associated with the 20\,cm radio source NVSS\,164843$+$800516. While a Galactic origin is not ruled out, this source is more likely to be extragalactic in nature.

 \subsection{XMMSL1\,J175542.2+624903}
This source was observed twice by both with {\it Swift} and {\it XMM-Newton} and shows X-ray variability. 
This source is coincident with 1RXS\,J175546.2$+$624927 discovered by {\it ROSAT}. It is listed in the Large Quasar Astrometric Catalogue \citep{Souchay} as being at redshift $z = 0.236$, and the same source is recorded in the optical identification of {\it ROSAT}-FSC sources \citep{Mickaelian}, {\it ROSAT} NEP X-ray source catalog \citep{Henry} and {\it ROSAT} North Ecliptic Pole Survey \citep{Gioia} classed as type 1 AGN. The RASS flux at the XRT position is a factor of 2 lower than during the XRT observation. The source type indicators presented in this paper are consistent with this classification.

 \subsection{XMMSL1\,J182707.5-465626}
Comparison with {\it XMM-Newton} flux and {\it ROSAT} flux limits shows this to be a strongly X-ray variable source. It is among the most variable of the detected sample when compared with {\it XMM-Newton} observations, decareasing by a factor of 11. 
From the X-ray to optical flux ratio and nIR colours this source could be an AGN, which would contradict the proper motion measurement of the optical counterpart.

\subsection{XMMSL1\,J185314.2-363057}
The X-ray to optical flux ratio places this source at the border between traditional AGN and low luminosity or less active galaxies however the optical counterpart displays proper motion and its nIR colours are typical of a  main sequence star of type M, strongly suggesting this source is Galactic.

\subsection{XMMSL1\,J185608.5-430320}
This source is detected with {\it Swift} BAT, with the highest significance (4$\sigma$) among the BAT-detected sources in this sample. One nIR counterpart is present within 15$''$ of the {\it XMM-Newton} Slew Survey position with $J \le 14$.

 \subsection{XMMSL1\,J211420.7+252419}
We extracted the X-ray flux seen by RASS at the new XRT enhanced X-ray position (Table \ref{tab:positions}) using the spectral shape measured by XRT (Table \ref{tab:spec}) and recover a detection. No variability is detected between the XRT and RASS observations 16.2 years apart while the flux decreased by a factor of $\sim$4 in the 3 years between the {\it XMM-Newton} and XRT observations. 

 \subsection{XMMSL1\,J215905.6-201604}
No single optical/nIR match was found in catalogue searches.: two optical sources lie within the XRT error circle, both at 14-15th magnitude. One lies 0.8$''$ from the XRT position and has a measured proper motion, while the other lies at 3$''$ with no PM measurement. X-ray variability is apparent: during the {\it Swift} observation the flux was (3$\pm$2)\% of that observed with {\it XMM-Newton} and $\le$20\% of that observed with {\it ROSAT}. A deeper investigation is needed to reveal the nature of this source.

\bsp

\label{lastpage}

\end{document}